\documentclass[letterpaper, submit]{AAS}			

\usepackage{bm}
\usepackage{amsmath}
\usepackage{amssymb}
\usepackage{graphicx}
\graphicspath{{Figures/}}
\usepackage{subcaption}
\usepackage{tikz}         
\usetikzlibrary{arrows.meta, calc}
\usepackage{booktabs}
\usepackage[colorlinks=true, pdfstartview=FitV, linkcolor=black, citecolor= black, urlcolor= black]{hyperref}
\usepackage{overcite}
\usepackage{footnpag}	
\usepackage{multirow} 
\usepackage{comment}

\PaperNumber{25-634}

\newcommand{\orcid}[1]{\href{https://orcid.org/#1}{\includegraphics[scale=.012]{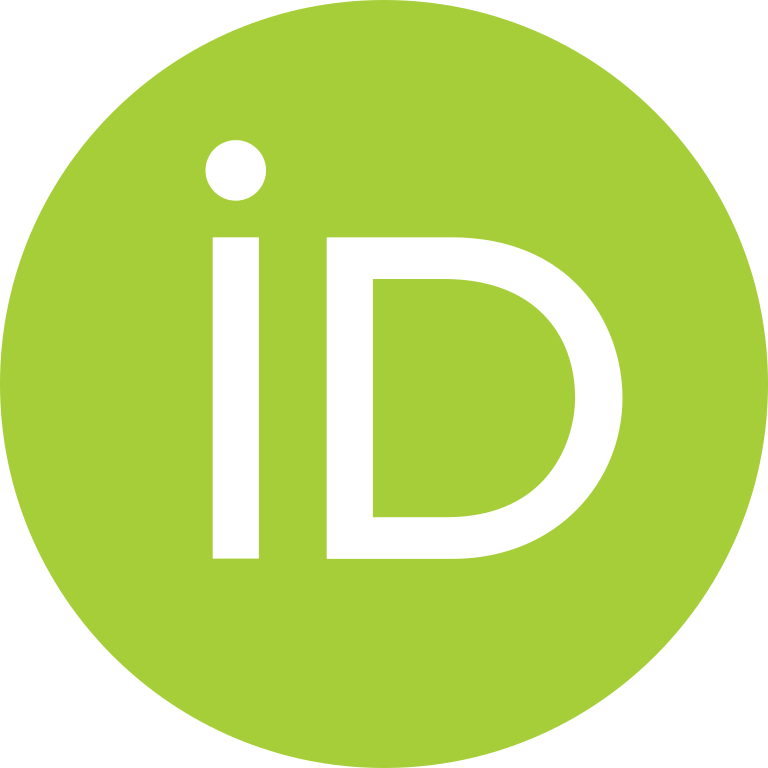}}}

\begin{document}

\title{Self-supervised diffusion model fine-tuning for costate initialization using Markov chain Monte Carlo
\thanks{
An abbreviated version of this paper was presented at the 2025 AAS/AIAA Astrodynamics Specialist Conference, Boston, MA, Aug. 2025, as Paper 25-634.
}
}

\author{Jannik Graebner \orcid{https://orcid.org/0009-0003-2497-124X}\thanks{PhD Student, Department of Mechanical and Aerospace Engineering, Princeton University, NJ, USA.}
\ and Ryne Beeson \orcid{0000-0003-2176-0976}\thanks{Assistant Professor, Department of Mechanical and Aerospace Engineering, Princeton University, NJ, USA.}
}

\maketitle{} 		

\begin{abstract}
Global search and optimization of long-duration, low-thrust spacecraft trajectories with the indirect method is challenging due to a complex solution space and the difficulty of generating good initial guesses for the costate variables.
This is particularly true in multibody environments.
Given data that reveals a partial Pareto optimal front, it is desirable to find a flexible manner in which the Pareto front can be completed and fronts for related trajectory problems can be found.
In this work we use conditional diffusion models to represent the distribution of candidate optimal trajectory solutions. 
We then introduce into this framework the novel approach of using Markov Chain Monte Carlo algorithms with self-supervised fine-tuning to achieve the aforementioned goals.
Specifically, a random walk Metropolis algorithm is employed to propose new data that can be used to fine-tune the diffusion model using a reward-weighted training based on efficient evaluations of constraint violations and missions objective functions. 
The framework removes the need for separate focused and often tedious data generation phases. 
Numerical experiments are presented for two problems demonstrating the ability to improve sample quality and explicitly target Pareto optimality based on the theory of Markov chains. 
The first problem does so for a transfer in the Jupiter-Europa circular restricted three-body problem, where the MCMC approach completes a partial Pareto front.  
The second problem demonstrates how a dense and superior Pareto front can be generated by the MCMC self-supervised fine-tuning method for a Saturn-Titan transfer starting from the Jupiter-Europa case versus a separate dedicated global search.  
\end{abstract}

\section{Introduction}
Long-duration, low-thrust spacecraft trajectory optimization is a complex optimal control problem involving multiple objectives and constraints. The solution space for this problem is characterized by funnel-like structures containing numerous local minima. Identifying pareto optimal solutions with respect to minimizing fuel consumption and time of flight is challenging and necessitates extensive exploration of this landscape.

Previous studies indicate that locally optimal solutions tend to cluster \cite{li2023amortizedglobalsearchtrajectory,beeson2024globalsearchoptimalspacecraft,Graebner.1032024,doi:10.2514/6.2012-4517,Yam.2011}  and global optimization algorithms such as Monotonic Basin Hopping (MBH) \cite{Wales.1997} exploit this clustering to expedite global searches.
MBH is based on simple sampling distributions and requires a substantial amount of manual tuning. 
Addressing these limitations, Li et al. \cite{li2023amortizedglobalsearchtrajectory} introduced the AmorGS framework, which utilizes deep generative models to represent complex conditional probability distributions focused on clusters of locally optimal solutions. This framework integrates state-of-the-art diffusion probabilistic models with numerical trajectory optimization solvers using a direct approach \cite{li2024diffusolvediffusionbasedsolvernonconvex}.
To further enforce the constraint statisfaction of the diffusion samples, Li et al. \cite{Li.412025} proposed constraint-aligned diffusion models, which incorporate a constraint violation loss during training.

Our previous work extended the AmorGS framework by integrating diffusion models with an indirect optimal control approach  \cite{Graebner.1132025}. 
While direct methods are frequently employed due to their simplicity and better convergence properties, indirect methods offer advantages such as reduced dimensionality in the control space and inherent satisfaction of first-order necessary optimality conditions \cite{Betts.1998}. 
However, these methods pose the significant challenge of identifying suitable initial guesses for non-intuitive costate variables. 
Our previous data-driven framework tackled this by learning a conditional probability distribution of initial-time costates for low-thrust spacecraft transfers. 
Despite its effectiveness, this approach had three primary limitations:
\begin{enumerate}
    \item Extensive upfront generation of training data is required.
    \item Its applicability, in terms of quality of samples, is likely restricted to the specific transfer scenario on which it was trained, allowing variation only within conditioned parameters (i.e. interpolation with potentially limited generalization).
    \item The diffusion model has no information of the feasibility or objective value of the costate initialization it generates, even though this data is easily available and could improve sample quality and diversity.
\end{enumerate}
\begin{figure}[t!]
\centering\includegraphics[width=\textwidth]{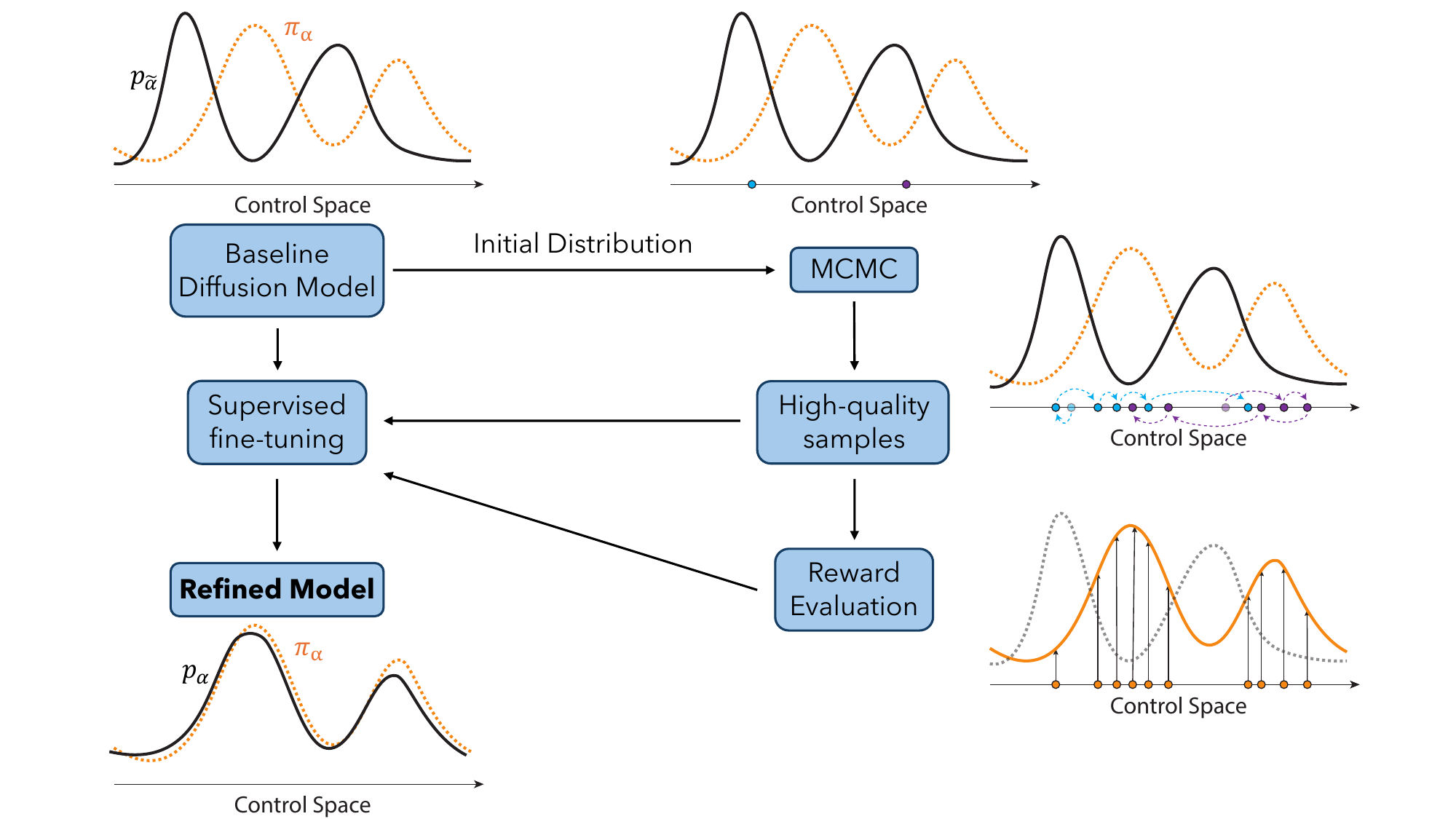}
	\caption{Simplified illustration of the proposed framework: Starting from a baseline diffusion model, with distribution $p_{\tilde{\alpha}}$ that is not aligned with the target distribution $\pi_\alpha$, MCMC and supervised-fine tuning are used to train a refined model which closely matches the target distribution.}
	\label{fig:MCMC_framework}
\end{figure}

In the current work, we address these limitations by proposing a new framework combining self-supervised diffusion model fine-tuning with a Markov Chain Monte Carlo (MCMC) approach, as illustrated in Figure \ref{fig:MCMC_framework}. 
The core idea is to start from a baseline diffusion model trained on a related problem (This often will be a different problem, but could be the same with a limited data set), generating samples from a distribution distinct yet structurally similar to our target distribution. 
We then employ an MCMC method to produce samples from the unknown distribution corresponding to our specific optimal control problem.
Through evaluation of a reward function, supervised fine-tuning \cite{Lee.2232023} is applied to refine the baseline model, thereby creating an improved model that closely matches the desired target distribution.

A significant advantage of this framework is that it theoretically does not require any solver-generated training data, since it is possible to start from any distribution and generate samples using MCMC.
From a practical standpoint, runtime and sample quality of the framework benefit significantly from starting the MCMC algorithm with samples from a baseline model, especially since previously trained models are available from our earlier work. 
Subsequently, fine-tuning to any (within a context of efficiency compromise) low-thrust transfer scenario becomes possible without additional offline data generation, effectively addressing limitations (1) and (2). 
Furthermore, limitation (3) is resolved by incorporating constraint violations and objective values into the reward evaluation step, efficiently computed through a preliminary screening algorithm, thereby enhancing sample quality.

The framework is designed to perform global search aimed at identifying a Pareto front of optimal solutions balancing fuel consumption and time of flight.
It thoroughly explores the solution space by combining global exploration through the baseline model's learned distribution with local perturbations via MCMC. 
A Gaussian random walk serves as the proposal density for the MCMC, enabling effective local exploration of funnel-like solution structures. 
The process of generating Pareto optimal solutions for a new problem can be facilitated by combining the MCMC steps with a homotopy scheme.
Additionally, the framework can be directly applied to improve the samples of a diffusion model already trained on the correct transfer but which may not yet fully capture the Pareto front.
We demonstrate the abilities of our framework by generating Pareto-optimal solutions for two low-thrust transfers in the Circular Restricted Three-Body Problem (CR3BP).

\section{Problem Formulation}
The following setup is similar to our previous work  \cite{Graebner.1132025} and we restate the problem setup for the convenience of the reader.
\subsection{Optimal Control Problem}
The goal of optimal control theory is to find an admissible  control input $\boldsymbol{u}(t)$ for a dynamical system,  that minimizes an objective function, while constraints on the state $\boldsymbol{x}(t)$ are satisfied \cite{Kirk.2004}.
In the general Bolza form, the problem is written as
\begin{align}
    \label{equation: cost functional}
    \min J(\boldsymbol{u}) \equiv \phi(\boldsymbol{x}(t_f), t_f) + \int_{t_0}^{t_f} \mathcal{L}(\boldsymbol{x}(t), \boldsymbol{u}(t), t) dt \quad 
    \textrm{subj. to Eqs. } 
    \eqref{equation: evolution differential equation}, \eqref{equation: initial and final boundary conditions}, \eqref{equation: path constraints} \ , 
\end{align}
where the objective function $J$ comprises a running cost $\mathcal{L}$ and a terminal cost $\phi$.
We consider a time interval $t\in[t_0,t_f]$, where in general the final time $t_f$ is treated as a free variable of the problem. 
The system dynamics are
\begin{align}
    \label{equation: evolution differential equation}
    \dot{\boldsymbol{x}}(t) = \boldsymbol{f}(\boldsymbol{x}(t), \boldsymbol{u}(t), t), \quad \forall t \in [t_0, t_f],
\end{align}
where the state trajectory satisfies the initial and terminal boundary conditions
\begin{align}
    \label{equation: initial and final boundary conditions}
    \boldsymbol{x}(t_0) = \boldsymbol{x}_0, \quad  \boldsymbol{\psi}\left[ \boldsymbol{x}(t_f), t_f \right] = \boldsymbol{0},
\end{align}
as well as a set of equality path constraints
\begin{align}
    \label{equation: path constraints}
    \boldsymbol{\xi}(\boldsymbol{x}(t), \boldsymbol{u}(t), t) = \boldsymbol{0}, \quad & \forall t \in [t_0, t_f].
\end{align}
The vector field $\boldsymbol{f}$ includes the natural dynamics of the system, as well as the perturbations induced by the control.

\subsection{Low Thrust Spacecraft Trajectory Optimization}
We aim to find low-thrust trajectories that simultaneously minimize fuel consumption and transfer time. 
A trajectory is labeled Pareto-optimal if no other feasible solution attains strictly lower values for both objectives.
We consider an engine with constant specific impulse $I{sp}$ and exhaust velocity $c=I_{sp}g_0$, where $g_0$ denotes standard gravity.
The state vector includes the spacecraft's position $\boldsymbol{r}\in\mathbb{R}^3$, velocity $\boldsymbol{v}\in\mathbb{R}^3$ and mass $m\in\mathbb{R}$:
\begin{equation}
\label{Equation: dynamics}
\boldsymbol{x}=\begin{pmatrix}
	\boldsymbol{r} \\
	\boldsymbol{v} \\
	m
	\end{pmatrix},
	\in \mathbb{R}^{7} \quad \text{and} \quad
    \dot{\boldsymbol{x}} = \boldsymbol{f}(\boldsymbol{x},\boldsymbol{u}) =
	\begin{pmatrix}
	\dot{\boldsymbol{r}} \\
	\dot{\boldsymbol{v}} \\
	\dot{m}
	\end{pmatrix}
	=
	\begin{pmatrix}
	\boldsymbol{v} \\
	\boldsymbol{g}(\boldsymbol{r},\boldsymbol{v}) + \frac{T}{m}\hat{\boldsymbol{u}} \\
	-\frac{T}{c}
	\end{pmatrix},
\end{equation}
are the equations of motion.
The natural system dynamics are described by the vector field $\boldsymbol{g}$.
A throttle variable $\sigma$ provides control over the thrust magnitude $T=\sigma T_{max}$, where $T_{max}$ is the maximum engine thrust.
To enforce the bounds $\sigma \in [0,1]$ as an equality constraint we introduce an additional slack variable $\zeta$ with $\sigma=\sin^2 \zeta$.
Combined with the unit thrust vector $\hat{\boldsymbol{u}}\in\mathbb{R}^3$, these form the control vector $\boldsymbol{u}=(\hat{\boldsymbol{u}}^\top,T,\zeta)^\top$. 
The set of admissible controls is encoded as two equality path constraints:
\begin{equation}
    \boldsymbol{\xi}(\boldsymbol{x}(t), \boldsymbol{u}(t), t) = 
    \begin{pmatrix}
	\hat{\boldsymbol{u}}^\top\hat{\boldsymbol{u}} - 1 \\
    T-T_{max}\sin^2\zeta
    \end{pmatrix} = \boldsymbol{0}.
\end{equation}
We consider a fixed initial state, as well as a fixed terminal position and velocity:
\begin{equation}
\label{Equation: Boundary conditions trajectory}
	\boldsymbol{x}(t_0) = \boldsymbol{x}_0, \quad\boldsymbol{\Psi}(\boldsymbol{x}(t_f)) =
    \begin{pmatrix}
	\boldsymbol{r}(t_f)- \boldsymbol{r}_f\\
    \boldsymbol{v}(t_f)- \boldsymbol{v}_f
    \end{pmatrix} = \boldsymbol{0},
\end{equation}
and the final time is treated as an free variable.
\subsection{Primer Vector Theory}
Indirect optimal‑control techniques employ the calculus of variations to formulate the first‑order necessary conditions for optimality \cite{Liberzon.2012}.  
When this framework is specialized to spacecraft trajectory optimization, it leads to the well‑known Primer Vector theory \cite{Rutherfobd.1964}.
As the first step the costate vector $\boldsymbol{\lambda} =(\boldsymbol{\lambda}_r^T,\boldsymbol{\lambda}_v^T, \lambda_m)^T \in \mathbb{R}^{7}$, consisting of position, velocity and mass costates is introduced to formulate the Hamiltonian:
\begin{equation}
	\label{Equation: Problem Hamiltonian}
	H = \mathcal{L} + \boldsymbol{\lambda}^\top \boldsymbol{f}(\boldsymbol{x},\boldsymbol{u}) = \boldsymbol{\lambda}_r^\top \boldsymbol{v} + \boldsymbol{\lambda}_v^\top \left(\boldsymbol{g}(\boldsymbol{r},\boldsymbol{v}) + \frac{T}{m}\hat{\boldsymbol{u}}\right) - \lambda_m\frac{T}{c}.
\end{equation}
Based on Pontryagin's Minimum principle \cite{Pontryagin.2018}, we seek to choose our control, such that $H$ is minimized.
This leads to $\hat{\boldsymbol{u}} = -\boldsymbol{\lambda}_v/\lambda_v$ and we rewrite:
\begin{equation}
	\label{Equation: Hamiltonian without control}
	H = \boldsymbol{\lambda}_r^\top \boldsymbol{v} + \boldsymbol{\lambda}_v^\top \boldsymbol{g}(\boldsymbol{r},\boldsymbol{v}) - S\frac{T}{m},\quad S = \lambda_v + \lambda_m m / c,
\end{equation}
where a switching function $S$ is introduced.
The throttle is chosen according to the "bang-bang" control law \cite{Conway.2010}, summarized as:
\begin{equation}
	\label{Equation: Control Law}
	\hat{\boldsymbol{u}} = -\frac{\boldsymbol{\lambda_v}}{\lambda_v}, \quad
	\sigma = 
	\begin{cases}
	0 & \text{if } S < 0 \\
	1 & \text{if } S > 0 \\
	0 \leq \sigma \leq 1 & \text{if } S = 0.
	\end{cases}
\end{equation}
Evaluating the first order necessary conditions for optimality also yields the costate dynamics \cite{Liberzon.2012}:
\begin{equation}
\label{Equation: Costate equation of motion}
\dot{\boldsymbol{\lambda}} = \begin{pmatrix}
\dot{\boldsymbol{\lambda}}_r \\
\dot{\boldsymbol{\lambda}}_v \\
\dot{\lambda}_m
\end{pmatrix} = \begin{pmatrix}
- \boldsymbol{G}^\top \boldsymbol{\lambda}_v \\
-\boldsymbol{\lambda}_r - \boldsymbol{H}^\top \boldsymbol{\lambda}_v \\
-\lambda_v T / m^2
\end{pmatrix}, \quad\text{where}\quad
\mathbf{G} = \frac{\partial \mathbf{g}}{\partial \mathbf{r}} \quad \text{and} \quad \mathbf{H} = \frac{\partial \mathbf{g}}{\partial \mathbf{v}}.
\end{equation}
Combined with the state equations from Eq. \eqref{Equation: dynamics}, as well as Eq. \eqref{Equation: Boundary conditions trajectory} and Eq. \eqref{Equation: Control Law} this yields a two point boundary value problem.
To fix the degrees of freedom introduced by the free final mass and final time, we set $\lambda_m(t_f)=-1$ and implicitly target the transversality condition
\begin{equation}
    \label{Equation: lam_m=-k}
     H(\boldsymbol{x}(t_f), \boldsymbol{u}(t_f), \boldsymbol{\lambda}(t_f), t_f)=0.
\end{equation}
The resulting problem can be solved numerically. 
\subsection{Circular Restricted Three Body Problem}
The CR3BP is an astrodynamics model, frequently employed during preliminary mission design. 
It describes the motion of a body of negligible mass under the gravitational force of two larger bodies.
The larger bodies are referred to as the primary with mass $m_1$ and the secondary with mass $m_2$, where $m_1>m_2$.
A system of natural units is used for normalizing position, time and velocity variables: the distance unit (DU) corresponds to the distance between primary and secondary, the time unit (TU) is the orbital period of primary and secondary divided by $2\pi$ and the mass unit (MU) is $m_1+m_2$.
The only parameter of the system is $\mu=m_2/(m_1+m_2)$.
The dynamics are described in a rotating frame of reference, with primary and secondary on the $r_1$ axis at $r_1=\mu$ and $r_2=1-\mu$.
They are given by the vector field
\begin{equation}
    \label{Equation: CR3BP dynamical system}
    \ddot{\boldsymbol{r}} = 
    \begin{pmatrix}
        \ddot{r}_1 \\
        \ddot{r}_2 \\
        \ddot{r}_3 
    \end{pmatrix}
    = \boldsymbol{g}(\boldsymbol{r},\boldsymbol{v}) =
    \begin{pmatrix}
        2v_2 + r_1 - (1-\mu)\frac{r_1+\mu}{\rho_1^3} - \mu\frac{r_1-1+\mu}{\rho_2^3} \\
        -2v_1 + r_2 - (1-\mu)\frac{r_2}{\rho_1^3} - \mu\frac{r_2}{\rho_2^3} \\
        -(1-\mu)\frac{r_3}{\rho_1^3} - \mu\frac{r_3}{\rho_2^3}
    \end{pmatrix},
\end{equation}
where
$\rho_1= \sqrt{(r_1+\mu)^2+r_2^2+r_3^2}$ and $\rho_2=\sqrt{(r_1-1+\mu)^2+r_2^2+r_3^2}$ are the distances to primary and secondary respectively.

\section{Methodology} 

\subsection{Diffusion Models} 
Recently, diffusion models have emerged as the new state-of-the-art class of deep generative models.
Inspired by principles from non-equilibrium thermodynamics, diffusion models were first introduced to machine learning by Sohl-Dickstein \cite{SohlDickstein.3122015}, with notable improvements by Song and Ermon \cite{Song.7122019} and Ho et al.\cite{Ho.6192020}.
Their application has since expanded into diverse domains, including reinforcement learning \cite{Wang.8122022,Ding.252024}, motion generation \cite{Tevet.9292022} and trajectory optimization \cite{li2023amortizedglobalsearchtrajectory}.

\begin{figure}[b!]
	\centering\includegraphics[width=\textwidth]{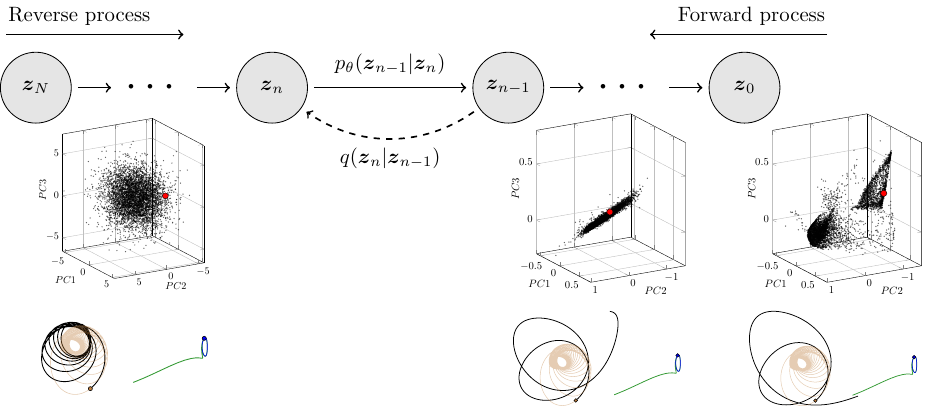}
	\caption{Visualization of the forward and reverse diffusion processes for a spacecraft trajectory optimization problem\cite{Graebner.1132025}. The distributions depict datasets of control vectors at various stages of the diffusion process with an example trajectory corresponding to the realization of the data-point marked in red.}
	\label{fig:diffusion_process}
\end{figure}

Diffusion models are a family of latent variable probabilistic models capable of modeling complex, high-dimensional distributions. 
They define a Markovian forward process that progressively perturbs data with Gaussian noise and subsequently train a neural network to approximate the corresponding reverse diffusion process.
This concept is visualized for the costate distribution of a indirect spacecraft trajectory optimization problem in Figure~\ref{fig:diffusion_process}.
The diffusion process can thus be decomposed into three main components: the forward process, the reverse process, and the sampling procedure. 
Below, these components are briefly explained, following the denoising diffusion probabilistic model formulation of Ho et al. \cite{Ho.6192020}.

\subsubsection{Forward Process}
Starting from training data sampled from a unknown distribution $\boldsymbol{z}_0 \sim q(\boldsymbol{z})$, the forward diffusion process is a discrete approximation of the continuous-time Markov process given by a stochastic differential equation.
By selecting $N$ discrete integration points, each forward step is modeled as sampling from a normal distribution:
\begin{align} \label{eq: forward diffusion process step}
  q(\boldsymbol{z}_n | \boldsymbol{z}_{n-1}) = \mathcal{N}(\boldsymbol{z}_n; \sqrt{1 - \beta_n}\boldsymbol{z}_{n-1}, \beta_n \boldsymbol{I}),\quad
  q(\boldsymbol{z}_{1:N} | \boldsymbol{z}_{0}) = \prod_{n=1}^N q(\boldsymbol{z}_n | \boldsymbol{z}_{n-1}),
\end{align}
where $\boldsymbol{I}$ is the identity matrix. 
A cosine-based variance schedule $\{ \beta_n \in (0, 1) \}_{n=1}^N$ controls the rate at which information is diluted.
 For practical training purposes, this process is reparameterized as \mbox{$\boldsymbol{z}_n =\sqrt{\overline{\alpha}_n} \boldsymbol{z}_0 + \sqrt{1 - \overline{\alpha}_n} \boldsymbol{\epsilon}$},
where $\boldsymbol{\epsilon} =  \mathcal{N}(\mathbf{0},\boldsymbol{I})$, $\alpha_n = 1 - \beta_n$ and $\overline{\alpha}_n = \prod_{i=1}^n \alpha_i$.

\subsubsection{Reverse Process}
The reverse process begins by sampling from a standard normal distribution $\boldsymbol{z}_N \sim  \mathcal{N}(\boldsymbol{0},\boldsymbol{I})$. 
Provided a sufficiently small step size, each step in the reverse process can also approximated by a Gaussian, with a mean predicted by a neural network:
\begin{align} \label{eq: reverse diffusion process}
	p_{\theta}(\boldsymbol{z}_{n-1} | \boldsymbol{z}_n) = \mathcal{N}(\boldsymbol{z}_{n-1}; \boldsymbol{\mu}_{\theta}(\boldsymbol{z}_n, n), \tilde{\beta}_n\boldsymbol{I}),
\end{align}
where $\theta$ denotes the network parameters and $\tilde{\beta}_n = (1 - \overline{\alpha}_{n-1})/(1 - \overline{\alpha}_n)\cdot \beta_n $.
The training objective maximizes the expectation of $p_\theta(\boldsymbol{z}_0)$ which equivalently involves minimizing a variational bound on the negative log-likelihood. 
After simplifications, this yields a simplified loss function:
\begin{equation} \label{eq: simplified loss}
L^{\text{simple}}(\theta) 
= \mathbb{E}_{P} 
\left[ \left\| \boldsymbol{\epsilon}_n - \boldsymbol{\epsilon}_{\theta}(\boldsymbol{z}_n, n) \right\|^2 \right],
\end{equation}
where $(\boldsymbol{z}_0, \boldsymbol{\epsilon}_n, n) \sim P \equiv \mathcal{D}\times\mathcal{N}(0, \mathbf{I})\times \text{Uniform}(\{1, \dots, N\})$. $P$ denotes the joint distribution over the data sample 
$\boldsymbol{z}_0$ (drawn from the training dataset $\mathcal{D}$), the Gaussian noise 
$\boldsymbol{\epsilon}_n$, and random selection of the 
time step index $n$ from a uniform distribution. 
This simplified loss is used to train a neural network based on the U-Net architecture, originally introduced by Ronneberger et al.\ \cite{Ronneberger.5182015}, employing stochastic gradient descent.
\subsubsection{Sampling}
Once training is completed, samples are generated through a sequential sampling loop from $n=N$ to $n=1$.
One step is given by:
\begin{equation} \label{eq: sampling}
	\boldsymbol{z}_{n-1} = \frac{1}{\sqrt{\alpha_n}} \left( \boldsymbol{z}_n - \frac{1 - \alpha_n}{\sqrt{1 - \alpha_n}} \boldsymbol{\epsilon}_{\theta} \right) + \sqrt{\tilde{\beta}_ n}\boldsymbol{y}(n) \quad
	\text{ where } \quad \boldsymbol{y}(n) 
	\begin{cases}
		=\boldsymbol{0} & \text{for } n = 1 \\
		\sim \mathcal{N}(\boldsymbol{0}, \boldsymbol{I}) & \text{otherwise.}
	\end{cases}
\end{equation}
No noise is added in the last step to obtain the final sample $\boldsymbol{z}_0$.

\subsection{Supervised Fine-tuning}
For large language models and text-to-image models, previous studies have demonstrated that generative model outputs can be improved by fine-tuning using human feedback \cite{Ouyang.342022,Lee.2232023,Ziegler.9182019}.
Typically, this involves training a separate reward model based on human evaluations of the generated outputs. 
This reward model is then used to assess the outputs of a  pre-trained baseline model.
The resulting outputs and their associated reward scores are subsequently used for model fine-tuning through reward-weighted likelihood maximization \cite{Lee.2232023}.

We adapt this approach for fine-tuning diffusion models designed to generate initial costates for trajectory optimization problems.
In this context, the reward function $R(\boldsymbol{z})$ can be directly computed as detailed in the subsequent section, thus eliminating the need to train an additional reward model.
The model parameters are updated by minimizing the loss function \cite{Lee.2232023}
\begin{equation}
L(\theta) = \mathbb{E}_{P^{\text{new}}}  \Big[ R(\boldsymbol{z}_0)\,\Big\| \boldsymbol{\epsilon}_n - \boldsymbol{\epsilon}_{\theta}(\boldsymbol{z}_n, n) \Big\|^2 \Big]
+\omega\,\mathbb{E}_{P^{\text{base}}} \Big[ \Big\| \boldsymbol{\epsilon}_n - \boldsymbol{\epsilon}_{\theta}(\boldsymbol{z}_n, n) \Big\|^2 \Big]
\label{eq:updated loss}
\end{equation}
where $P^{\text{base}}$ samples from the dataset used to train the baseline model ($\mathcal{D}^{\text{base}}$), and $P^{\text{new}}$ represents newly acquired data ($\mathcal{D}^{\text{new}}$).
Both $P^{\text{base}}$ and $P^{\text{new}}$ are product distributions in the same sense as $P$. 
The loss term associated with new data is scaled so that each sample’s contribution to training is modulated by its reward value.
We incorporate the baseline training loss with a penalty parameter $\omega\in\mathbb{R}_{\geq0}$, as image generation studies have demonstrated that this approach helps mitigate overfitting \cite{Ouyang.342022}. The penalty parameter is chosen as the mean reward over the new samples,
$\omega=\mathbb{E}_{\mathcal{D}^{\text{new}}}[R(\boldsymbol{z}_0)]$
to ensure a similar scaling. 
We introduce the parameter $\eta = |\mathcal{D}^{\text{base}}_{\text{sub}}|/|\mathcal{D}^{\text{new}}|$, where $|\mathcal{D}^{\text{base}}_{\text{sub}}|$ is the number of baseline samples included and $|\mathcal{D}^{\text{new}}|$ the number of new samples.
This ratio is used in the results section to control the contribution of baseline data during training. 


\subsection{Reward Evaluation}
\label{sec:reward_eval}
For reward evaluation, we use a preliminary screening algorithm, which is described in the author's previous work \cite{Graebner.1132025} and was inspired by the work of Russell \cite{Russell.2007}.
The algorithm relies on \texttt{pydylan}, the Python interface of the astrodynamics software package Dynamically Leveraged (N) Multibody Trajectory Optimization (DyLAN) \cite{Beeson.Aug.2022}.
The decision vector of the indirect method in DyLAN is:
\begin{equation}
    \label{Equation: decision vector indirect}
    \overline{\boldsymbol{u}} = (\tau_s,\tau_i,\tau_f,\boldsymbol{\lambda}_{r,0}^\top,\boldsymbol{\lambda}_{v,0}^\top,{\lambda}_{m,0})^\top,
\end{equation}
where $\overline{\boldsymbol{u}}\in \mathbb{R}^{10}$ comprises the seven initial costates and three time variables.
The shooting time $\tau_s$ during which the actual maneuver happens is enclosed by an initial and final coast $\tau_i$ and $\tau_f$, during which the spacecraft’s thrusters are inactive.

For a given initial‑costate sample $\boldsymbol{\lambda}_0$, the preliminary screening algorithm simultaneously:
\begin{itemize}
    \item selects optimal values $\tau_s^{*}\in[0,\tau_{s,max}]$ and $\tau_f^{*} \in [0,\mathcal{T}_f]$ while fixing $\tau_i=0$,
    \item evaluates the constraint violation $e(\boldsymbol{\lambda}_0)$ (see Eq. \eqref{eq: constr violation}) at the terminal state and 
    \item determines the corresponding final mass $m_f(\boldsymbol{\lambda}_0)$.
\end{itemize}
Here, $\mathcal{T}_f$ denotes the orbital period of the target orbit.
Following a forward shooting approach, the costates only need to be propagated once under CR3BP dynamics for a prescribed maximum shooting time $\tau_{s,max}$.
The pair $\tau_s^*$ and $\tau_f^*$, is chosen so that the corresponding states on the transfer trajectory and on the target orbit attain the minimum Euclidean distance, found efficiently with a k‑dimensional (k-d) tree search \cite{Bentley.1975}.
The resulting constraint violation is
\begin{equation}
\label{eq: constr violation}
    e(\boldsymbol{\lambda}_0)=\left\Vert
    \begin{pmatrix}
    \boldsymbol{r}(\tau_s^*) - \boldsymbol{r}_f(\tau_f^*) \\
    \boldsymbol{v}(\tau_s^*) - \boldsymbol{v}_f(\tau_f^*)
    \end{pmatrix}\right\Vert_2,
\end{equation}
where $(\boldsymbol{r}_f^\top,\boldsymbol{v}_f^\top)$ is a state on the final orbit, parameterized by $\tau_f^*$.
The objective function chosen in this work combines the constraint violation with the two objectives:
\begin{equation}
\label{eq:objective function}
    J(\boldsymbol{\lambda}_0)=e(\boldsymbol{\lambda}_0)+\kappa_1 \left(\frac{\Delta m(\boldsymbol{\lambda}_0)}{m_0}+\kappa_2\tau_s (\boldsymbol{\lambda}_0)\right),
\end{equation}
where $\Delta m =m_0-m_f$ denotes the consumed fuel mass.
All three terms in the objective are computed efficiently through the preliminary screening algorithm.
The scaling parameter $\kappa_1$ controls the trade-off between feasibility and optimality, whereas $\kappa_2$ allows different weighting of 
the fuel consumption and time of flight objectives.
The reward $R$ in Eq. \eqref{eq:updated loss} is chosen based on this objective function, with their specific relation detailed in Eq. \eqref{eq: reward} in the results section.

\subsection{Markov Chain Monte Carlo} 
MCMC is a statistical method used to sample from complex probability distributions that are only known up to a normalization constant \cite{Metropolis.1953,Hastings.1970}.
Direct Monte Carlo sampling requires independent draws from the normalized target distribution, which is typically intractable in high-dimensional problems because the normalization integral cannot be evaluated.
MCMC circumvents this difficulty by constructing a Markov chain $(X_k)$ whose invariant distribution coincides with the target distribution \cite{Douc.2018}.

The most common class of MCMC algorithms is based on the Metropolis–Hastings scheme. 
Beginning with a sample from an initial distribution $X_0 \sim \nu$, a candidate $Y_{k+1}$ is drawn from a tractable proposal distribution $Y_{k+1} \sim q(\cdot ; X_k)$.
The proposal distribution along with an acceptance rule constitutes the transition probability of the Markov chain. 
In Metropolis-Hastings, the acceptance of a sample from the proposal is given with probability
\begin{equation}
    \label{eq: acceptance probability}
\alpha(X_k,Y_{k+1})=\text{min}\left(1,\frac{\pi(Y_{k+1})q(Y_{k+1}; X_k)}{\pi(X_k)q(X_k; Y_{k+1})}\right),
\end{equation}
where $\pi$ is the unnormalized target density.
Accepted samples define the next state of the chain, while rejected ones leave the state unchanged. 
Under mild regularity conditions, the resulting chain converges asymptotically to the target distribution \cite{10.1214/aos/1176325750}.
Therefore running the chain for a long enough duration results in Monte Carlo samples from the desired target distribution. 

In this work we employ the random walk Metropolis algorithm, a specialization of Metropolis–Hastings where the proposal distribution is a symmetric distribution. 
In particular, we use a Gaussian distribution with mean given by the current state of the chain,
\begin{equation}
    \label{eq: proposal distribution}
    q(\tilde{\boldsymbol{\lambda}}^{k+1}; \boldsymbol{\lambda}^k) 
    \equiv \mathcal{N}(\tilde{\boldsymbol{\lambda}}^{k+1}; \boldsymbol{\lambda}^k,\boldsymbol{\Sigma}_{\lambda}).
\end{equation}
Because the proposal is symmetric, the acceptance probability simplifies to
\begin{equation}
    \label{eq: acceptance probability simple}
\alpha(X_k,Y_{k+1})=\text{min}\left(1,\frac{\pi(Y_{k+1})}{\pi(X_k)}\right).
\end{equation}
This formulation is straightforward to implement and guarantees that samples that move to a higher density region of the target distribution are always accepted.
Additionally, samples from a lower probability region may be accepted at a low rate, and provide an avenue for continued exploration by the Markov chain. 
In practice, chains are run for a fixed number of iterations, discarding an initial burn-in phase before retaining samples for analysis \cite{Brooks.1998}.
The burn-in phase reflects the fact that the chain may require many iterations before it provides samples close to that of the target distribution. 
The more similar the initial distribution is to the target distribution, and the more effective the MCMC algorithm, the shorter the required burn-in period.
Multiple independent chains can be executed in parallel to improve coverage of the state space \cite{CharlesJ.Geyer.1992}.

\subsection{Self-supervised Training Framework}
\begin{figure}[b!]
\centering\includegraphics[width=0.8\textwidth]{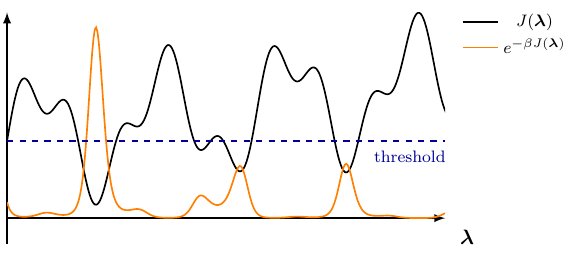}
	\caption{Reformulating optimization into sampling: minima of $J(\boldsymbol{\lambda})$ correspond to peaks of the unnormalized target density $\pi_{\alpha}(\boldsymbol{\lambda})\equiv\exp(-\beta J(\boldsymbol{\lambda}))$.}
	\label{fig: sampling function}
\end{figure}

Our goal is to sample from a target probability distribution that generates high-quality samples for the initial costates $\boldsymbol{\lambda}_0 \in \mathbb{R}^n$ of a low-thrust spacecraft transfer parameterized by problem parameters $\boldsymbol{\alpha}$.
To streamline notation, we henceforth omit the subscript “0” and denote the initial-time costate vector simply by $\boldsymbol{\lambda}$.
By defining the unnormalized target density as 
\begin{equation}
    \pi_{\alpha}(\boldsymbol{\lambda})\equiv\exp(-\beta J(\boldsymbol{\lambda})),
\end{equation}
we reformulate the optimization problem (minimization of $J(\boldsymbol{\lambda})$) as a sampling problem (drawing from $\pi_\alpha(\boldsymbol{\lambda})$).
Here $J(\boldsymbol{\lambda})$ is the objective function from Eq. \eqref{eq:objective function}.
Figure \ref{fig: sampling function} illustrates how the resulting high-density regions are centered around high-quality local minima of the objective function. 
The scaling parameter $\beta$ prevents the exponent from being too small and is chosen to yield practical acceptance rates in the range of $20-80\,\%$ in the random walk Metropolis algorithm. 
Our framework seeks to generate approximate samples from $\pi_\alpha(\boldsymbol{\lambda})$ by coupling a diffusion model with a random walk Metropolis algorithm.
The entire process is illustrated in Figure \ref{fig:MCMC_framework}.

\subsubsection{Baseline Diffusion Model} 
The initial distribution $\nu(\boldsymbol{\lambda})$ for the Markov chain is given through a baseline diffusion model. 
This model might have been trained on a different transfer with parameters $\Tilde{\boldsymbol{\alpha}}$, but ideally generates a density function $p_{\Tilde{\alpha}}(\boldsymbol{\lambda})$ which is similar to $\pi_{\alpha}(\boldsymbol{\lambda})$.
Starting with $N$ samples from the model, we launch $N$ independent Markov chains.
The state of a chain at iteration $k$ is denoted by $\boldsymbol{\lambda}^k$, with initial states drawn as $\boldsymbol{\lambda}^0 \sim \nu$.
\subsubsection{MCMC} 
Each chain evolves using the random walk Metropolis algorithm and proposes a new costate $\tilde{\boldsymbol{\lambda}}^{k+1}$ according to Eq. \eqref{eq: proposal distribution}.
This random-walk step ensures local exploration of the solution space around the generated samples and helps to escape local minima.
The proposal covariance matrix $\boldsymbol{\Sigma}$, which is symmetric and positive definite, is chosen to be diagonal with each entry dependent on the sensitivity of the problem with respect to the corresponding costate. 
We run each chain for $M$ iterations, discard the first $M_0$ samples as burn‑in, and retain the remaining $M-M_0$ states.  
Across all chains, this yields $(M-M_0)N$ approximate samples from the target distribution.
If the initial distribution differs significantly from the target, the MCMC algorithm is coupled with a homotopy scheme.
By incrementally adjusting the problem parameters bridging the two distributions, we introduce intermediate distributions that facilitate convergence.
The final states from all MCMC chains at one homotopy stage then serve as the initial samples for the subsequent stage.
\subsubsection{Reward Evaluation}
For each of these samples, which is very likely to be of high-quality, we evaluate the corresponding reward values.
The reward function $R$ in the loss term from Eq. \eqref{eq:updated loss} is chosen based on the objective function, with a slightly different scaling than the target density:
\begin{equation}
    \label{eq: reward}
    R(\boldsymbol{\lambda})\equiv a\exp(-b J(\boldsymbol{\lambda})).
\end{equation}
We include $a$ and $b$ as problem-specific parameters that ensure the reward function spans the range $[0.1,1]$, therefore attributing the best sample ten times the weight of the worst.
\subsubsection{Supervised fine-tuning}
The collected costate–reward pairs $\{(\boldsymbol{\lambda},R(\boldsymbol{\lambda})) \}$ are then used to fine-tune the baseline diffusion model. Training proceeds by reward-weighted likelihood maximization with the loss function of Eq. \eqref{eq:updated loss}, which biases the model toward generating samples with higher reward values.
Through this combination of diffusion models with MCMC in a self-supervised fine-tuning loop, we achieve an improved fit of the density $p_\alpha$ learned by the refined model to the target density $\pi_{\alpha}(\boldsymbol{\lambda})$.

\section{Results}
We present results for two representative problems illustrating different applications of the proposed framework.
First, we fine-tune a baseline diffusion model already trained to generate conditional samples for the correct transfer, aiming to improve sample quality and completion of the Pareto front. 
We then transition to tackling a more challenging scenario of generating costate samples for a new transfer with different parameters.
Both transfers are planar, so each sample comprises a four-dimensional costate vector. 
The samples generated by the MCMC algorithm, together with their associated rewards, are then used to fine-tune the baseline model (Problem 1) or to train a new model (Problem 2).
Other than the new reward-weighting term, the diffusion model’s architecture and hyperparameters are identical to those detailed in our earlier work \cite{Graebner.1132025}. 
In both cases, training takes under 15 minutes, and sampling 100,000 costate vectors takes under 20 minutes on a single GPU.

\subsection{Problem 1: Model Improvement for Europa DRO Transfer}
\begin{figure}[t!]
  \centering
  \begin{minipage}[t]{0.50\textwidth}
    \vspace{0pt}
    \centering
    \includegraphics[width=\linewidth]{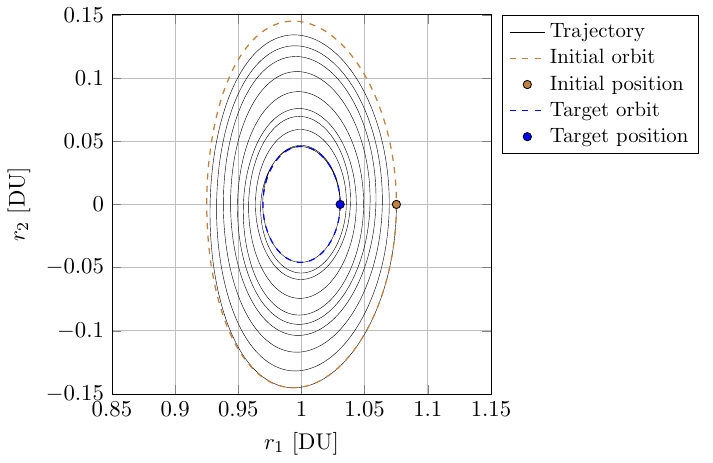}
    \caption{Example trajectory Europa DRO transfer.}
    \label{fig:ex_traj_0.95}
  \end{minipage}%
  \hfill
  \begin{minipage}[t]{0.45\textwidth}
    \vspace{0pt}
    \centering
    \captionof{table}{MCMC parameters Europa DRO transfer.}
    \begin{tabular}{@{}ll@{}}
      \toprule
      \textbf{Input Parameters} & \textbf{Value}\\
      \midrule
      Objective scaling $\kappa$ & $1.0$\\
      Shooting time scaling $\kappa_2$ & $5\times10^{-6}$\\
      Proposal standard deviations $\boldsymbol{\sigma}$ & $0.01 \times \boldsymbol{\sigma}_{\text{init}}$ \\
      Scaling factor $\beta$ & $10{,}000$\\
      Number of chains $N$ & $1{,}400$\\
      Number of iterations $M$ & $6{,}000$\\
      Burn-in iterations $M_0$ & $5{,}950$\\
      \midrule
      \multicolumn{2}{@{}l}{\textbf{Resulting Metrics}}\\
      Total function evaluations $MN$ & $8,400,000 $\\
      Number of final samples & $34,757$\\
      Runtime [CPU-hours] & 405\\
      \bottomrule
\end{tabular}
    \label{tab:mcmc_params_europa}
  \end{minipage}
\end{figure}

\begin{figure}[t!]
    \centering
    \begin{tikzpicture}[every node/.style={inner sep=0pt}]
        \newlength{\colw}      
        \setlength{\colw}{0.33\textwidth}   
        \def\vsep{5.3cm}       

        \node (L1) at (-5.8,   0)   {\includegraphics[width=0.95\colw]{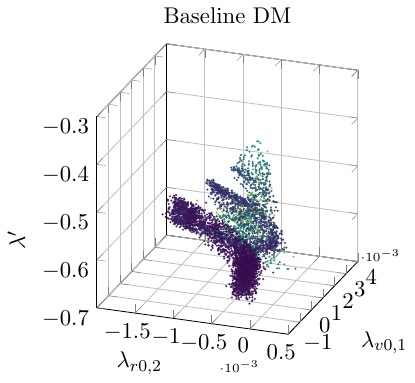}};
        \node (L2) at (-5.8,-\vsep) {\includegraphics[width=0.95\colw]{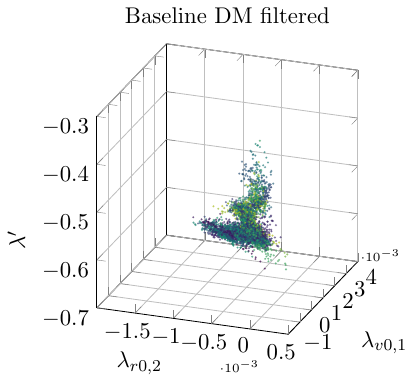}};

        \node (C)  at ( -0.7, -0.5*\vsep)
                       {\includegraphics[width=1.08\colw]{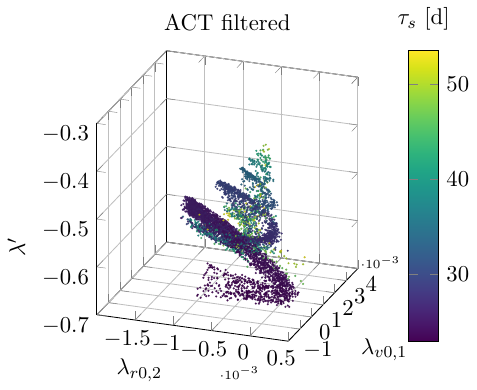}};

        \node (R1) at ( 4.4,   0)   {\includegraphics[width=0.95\colw]{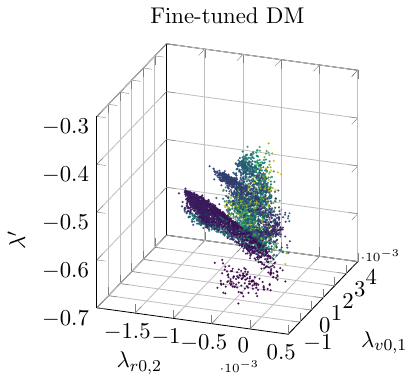}};
        \node (R2) at ( 4.4,-\vsep) {\includegraphics[width=0.95\colw]{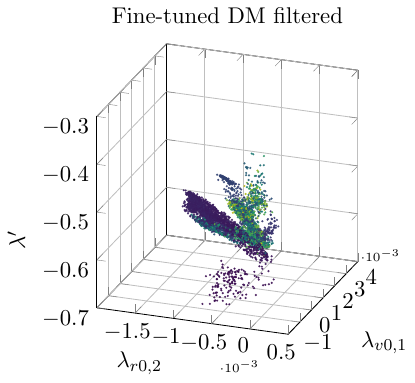}};
        \draw[->, line width=0.8mm, >=stealth] ($(L1.south)$) -- (L2.north);
        \draw[->, line width=0.8mm, >=stealth] ($(R1.south)$) -- (R2.north);
        \draw[->, line width=0.8mm, >=stealth]
              ($(L1.north east)+(2mm,-10mm)$)              
              to[out=20,in=160]                         
              node[midway, above, yshift=1mm, font=\scriptsize] {Supervised fine-tuning}              
              ($(R1.north west)+(-2mm,-10mm)$); 
    \end{tikzpicture}

    \caption{Comparison of the structure in costate space before and after supervised finetuning.  
             Left: diffusion model (DM) samples and their feasible subset;  
             Center: separately generated solutions using ACT;  
             Right: fine-tuned DM samples and their feasible subset.}
    \label{fig:costate-structure-finetune}
\end{figure}

\begin{figure}[t!]
  \centering
  \begin{tikzpicture}[every node/.style={inner sep=0pt}]
    \node (L) at (-1,0)
          {\includegraphics[height=6.0cm]{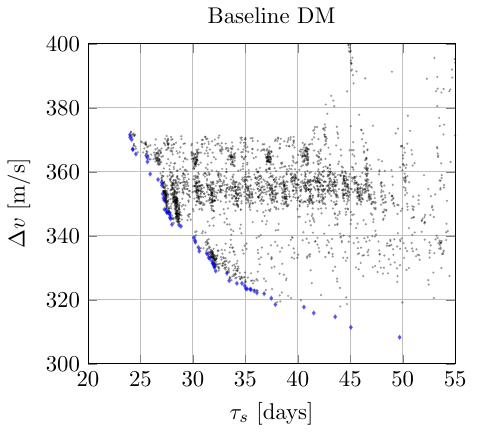}};
    \node (R) at (7.0,0)  
          {\includegraphics[height=6.0cm]{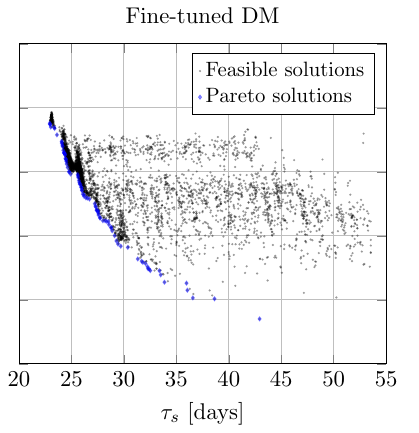}};

    \draw[->, line width=0.8mm, >=stealth]
          ($(L.north east)+(-3mm,0)$)   
          to[out=10,in=170]            
          node[midway, above, yshift=1mm, font=\small]
               {Supervised fine-tuning}
          ($(R.north west)+(3mm,0)$); 
  \end{tikzpicture}

  \caption{$5,000$ feasible samples from the baseline and the fine-tuned model in the $\Delta v$-$\tau_s$ plane.}
  \label{fig:eur_pareto}
\end{figure}

\begin{table}[b!]
\centering
\caption{Problem parameters Europa DRO transfer.} 
\setlength{\tabcolsep}{8pt} 
\begin{tabular}{lclc} 
\toprule
\multicolumn{2}{l}{\textbf{Trajectory parameters}} & \multicolumn{2}{c}{} \\
\midrule
\multicolumn{2}{l}{Initial state $[\boldsymbol{r}_0^T,\boldsymbol{v}_0^T]$ [NU]} & \multicolumn{2}{c}{$[1.0752, 0.0, 0.0, 0.0, -0.1499, 0.0]$}\\
\multicolumn{2}{l}{Terminal state $[\boldsymbol{r}_f^T,\boldsymbol{v}_f^T]$ [NU]} & \multicolumn{2}{c}{$[1.0306, 0.0, 0.0, 0.0, -0.0727, 0.0]$}\\
\multicolumn{2}{l}{Orbital period target DRO $\mathcal{T}_f$ [TU]} & \multicolumn{2}{c}{$4.1055$}\\
\multicolumn{2}{l}{Max. shooting time $\tau_{s,.max}$ [TU]} & \multicolumn{2}{c}{$90$}\\
\midrule
\multicolumn{2}{l}{\textbf{Spacecraft parameters}} & \multicolumn{2}{l}{\textbf{Natural units (Jupiter-Europa)}} \\
\midrule
Initial mass $m_0$ [kg] & 25,000 & Distance unit [km] & 670,900 \\
Fuel mass [kg] & 15,000 & Time unit [s] & 48,822.76 \\
Dry mass [kg] & 10,000 & Mass unit [kg] & $1.898 \times 10^{27}$ \\
Specific impulse $I_{sp}$ [s] & 7,365 & Mass parameter $\mu_{JE}$ & $2.528 \times 10^{-5}$ \\
Thrust magnitude $T_{max}$ [N] & $4.735$ & & \\
\bottomrule
\end{tabular}
\label{tab:problem parameters Europa DRO}
\end{table}
For this problem we apply the presented framework to improve diffusion model sampling results, obtained in our previous work \cite{Graebner.1132025}.
The objective is to generate Pareto-optimal solutions for a spacecraft transfer in the Jupiter-Europa CR3BP system. 
This problem derives from the Jupiter Icy Moons Orbiter (JIMO) mission concept, which was canceled in 2005 and involves using a low-thrust propulsion system to transfer from a high Distant Retrograde Orbit (DRO) around Europa to a lower DRO. 
An example trajectory is shown in Figure \ref{fig:ex_traj_0.95}.
The detailed trajectory and spacecraft parameters as well as the natural units of the Jupiter-Europa system are listed in Table \ref{tab:problem parameters Europa DRO}.
In our prior study \cite{Graebner.1132025} we solved this problem for varying values of the thrust magnitude $T_{max}$, which served as a conditional variable for diffusion model training. 
Although the model correctly predicted the structure in the costate space for unseen values of $T_{max}$ and accelerated the solution generation for these problems, its accuracy degraded at $T_{max}=4.735\,\mathrm{N}$.
The reason for this is a slightly wrong prediction of the costate structure, as shown on the left of Figure \ref{fig:costate-structure-finetune}, in comparison to separately generated data using Adjoint Control Transformations (ACT)\cite{Dixon.1981}, shown in the center of Figure \ref{fig:costate-structure-finetune}\footnote{This data was separately generated by combining ACT with a preliminary screening algorithm and was used for comparison and benchmarking the model in our previous work where the method is described in more detail \cite{Graebner.1132025}.}.
This results in a low feasibility rate for diffusion model samples ($<1\,\%$) and a sparse Pareto front in the $\Delta v$-$\tau_s$-plane, visualized on the left of Figure \ref{fig:eur_pareto}.

\begin{figure}[b!]
	\centering\includegraphics[width=\textwidth]{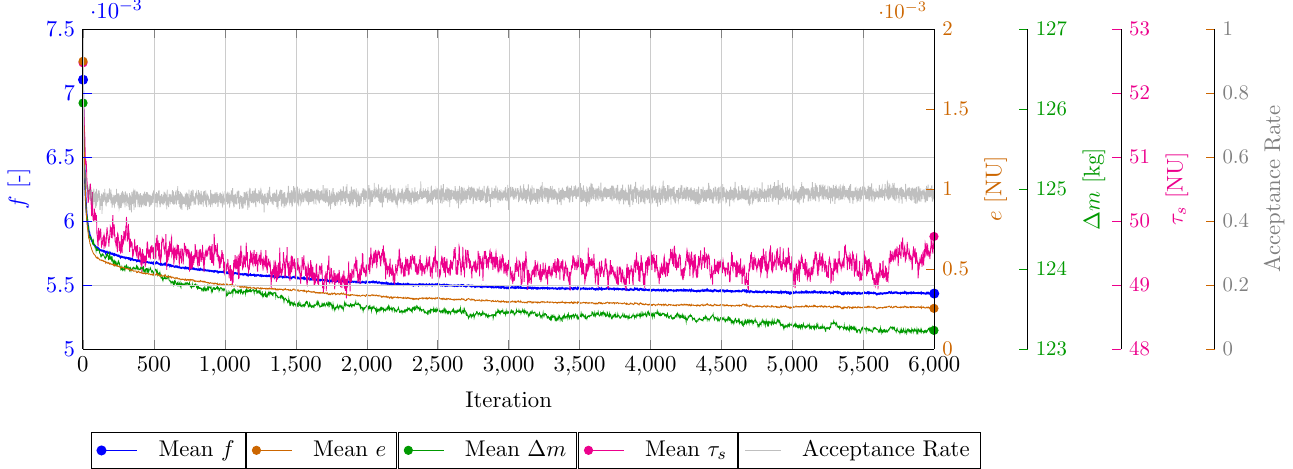}
	\caption{Mean values of the objective function and its three components during MCMC for the Europa DRO problem.}
	\label{fig:MCMC_iter_europa}
\end{figure}

We address this deficit by (i) running a random-walk Metropolis MCMC and (ii) fine-tuning the diffusion model with reward-weighted likelihood maximization.
Samples from the initial distribution $p_{\tilde{\alpha}}$ of the random-walk Metropolis algorithm are generated by drawing from the conditional distribution learned by the baseline model given $T_{max}=4.735\,\mathrm{N}$.
Key MCMC parameters are listed in Table \ref{tab:mcmc_params_europa}.
The most important and difficult parameter choice is the diagonal covariance matrix of the proposal distribution from Eq. \eqref{eq: proposal distribution}.
A naive approach of using the same value for each direction did not perform well for this problem, as the problem scaling depends heavily on the costate direction.  
If this fixed value is chosen too small, there is not enough mixing in some directions, whereas larger values lead to diminishing acceptance rates.
Another method that was tested includes conducting a pilot run with fixed covariance in each direction and choosing a multiple of the empirical covariance of the generated samples as the new proposal covariance. 
While this method achieves improved performance and decreases the mean objective value, the additional pilot run increases the computational cost. 
The best performance is achieved by choosing the proposal covariance $\boldsymbol{\Sigma}=\text{Diag}(\boldsymbol{\sigma})$ based on the empirical standard deviation of the initial samples $\boldsymbol{\sigma}_{\text{init}}$.
Here $\boldsymbol{\sigma}$ is a vector consisting of the  standard deviation $\sigma_i$ for each direction $i$.
The standard deviation multiplier $0.01$ in Table \ref{tab:mcmc_params_europa} was selected as, in combination with the scaling factor $\beta$, it leads to a steady decrease in the objective function and an acceptance rate around $50\,\%$ (see Figure \ref{fig:MCMC_iter_europa}).

We run the MCMC algorithm for a total of 405 CPU-hours distributed across 28 cores.
The computational cost associated with this long runtime is currently still high and we will aim to improve the efficiency of the framework in future work.
After discarding the burn-in samples and selecting the best $50\,\%$ (based on $J$), a total of $17,000$ unique samples are used for model fine-tuning.
In comparison $27,000$ samples were used for training the baseline model at a single thrust level (with the full baseline dataset comprising $270,000$ samples across all thrust levels).
For the initial studies presented we only use new data for fine-tuning ($\eta=0.0$), but an additional study for different data ratios is included at the end of the section.

Starting from a set of low-quality initial costates, the MCMC sampler progressively improves sample quality.
Figure \ref{fig:MCMC_iter_europa} shows a monotonic decline in the mean objective $J(\boldsymbol{\lambda})$, constraint violation $e(\boldsymbol{\lambda})$, fuel consumption $\Delta m (\boldsymbol{\lambda})$ and shooting time $\tau_s$.
The objective curve flattens toward the end of the run, indicating that additional iterations are unlikely to offer a substantial improvement in accuracy.
The choice $\beta=10{,}000$ results in an acceptance rate that fluctuates around $50\,\%$, which provides a reasonable trade-off between sufficient mixing to explore new solutions and avoiding excessive acceptance of inferior samples.

Figure \ref{fig:costate-structure-finetune} compares the costate distributions:
\begin{itemize}
    \item Left: Baseline diffusion-model (DM) samples and their feasible subset (Baseline DM filtered);
    \item Center: reference solutions generated via adjoint control transformations (ACT);
    \item Right: Fine-tuned DM samples and their feasible subset (Fine-tuned DM filtered).
\end{itemize}
The feasible subset is selected based on the constraint violation with $e<10^{-4}$ NU.
While this is a fairly coarse tolerance from a mission design perspective, we emphasize that this work focuses on low-fidelity global search with tight tolerances not being a primary focus. 
To visualize the four-dimensional costate vector in three dimensions, we apply the linear transformation
\begin{equation}
    \label{eq: lambda_trans}
    \lambda' = 0.5895\lambda_{r0,1}+0.8075\lambda_{v0,2}.
\end{equation} 
The reason for this variable transformation is a linear relationship between $\lambda_{r0,1}$ and $\lambda_{v0,2}$ identified in our previous work \cite{Graebner.1032024}.
After fine-tuning, the distribution learned by the model aligns more closely with the hypersurface structure of the ACT solutions shown in Figure \ref{fig:costate-structure-finetune}.

In particular, after fine-tuning, the model also partially captures the lowest branch of samples, which was not targeted by the baseline model. 
This improvement is also observed in the filtered samples, where the fine-tuning prevents the collapse to a smaller region of the costate space evident in the baseline model.

The fine-tuning process reveals a smooth and full front of Pareto optimal solutions in the $\Delta v$-$\tau_s$ plane, shown in Figure \ref{fig:eur_pareto}.
The wave-structure of the Pareto front closely matches results from Russell \cite{Russell.2007} for the same problem with different thrust magnitude.
The significant improvement from the baseline model demonstrates how this framework nudges samples towards Pareto optimality.
As both the fuel consumption and the time of flight are included in the objective function and the target density, states with lower objective values are assigned exponentially higher probability during MCMC, and the reward-weighted loss further amplifies this preference during training.
For shooting times above $\tau_s=30$ days the Pareto front becomes more sparse, revealing some limitations of the framework. 
With not enough samples from the baseline model being located in that region of the solution space, the MCMC algorithm fails to explore long-duration solutions effectively for the given number of iterations.
Given a sufficiently long runtime, the algorithm would presumably be able to populate that region as well, although the timescale at which this occurs is unclear.

Tuning the weighting parameters $\kappa_1$ and $\kappa_2$ in the objective function from Eq. \eqref{eq:objective function} allows for some control over the targeted region in the $\Delta v$-$\tau_s$ landscape. 
Table \ref{tab:kappa-gamma-comp} shows a comparison of averaged metrics for different parameter combinations, with the baseline model as a reference.
\begin{table}[b!]
  \centering
  \caption{Performance metrics based on samples from the fine-tuned diffusion model for various $(\kappa_1,\kappa_2)$ combinations in comparison to samples from the baseline model. All metrics are computed based on $100{,}000$ samples with feasibility tolerance $e<10^{-4}$ NU. The center column ($\kappa_1=1.0, \kappa_2=5\times10^{-6}$, RW) correspond to the selected model with samples shown in Figure \ref{fig:MCMC_iter_europa} and \ref{fig:eur_pareto}. For this case results with and without reward weighting (RW) are shown side by side (all other fine-tuned models use reward weighting).}
  \begingroup
  \setlength{\tabcolsep}{3pt}
  \renewcommand{\arraystretch}{1.05}
  \small
  \begin{tabular}{lccccccc}
    \toprule
    &  
    & \multicolumn{6}{c}{Fine-tuned Model} \\[-1.0ex]
    \cmidrule(lr){3-8}
    & \shortstack{Baseline\\Model}
    & \shortstack{$\kappa_1 = 0.5$\\$\kappa_2 = 0.0$}
    & \shortstack{$\kappa_1 = 1.0$\\$\kappa_2 = 1\times10^{-6}$}
    & \multicolumn{2}{c}{\shortstack{$\boldsymbol{\kappa_1 = 1.0}$\\$\boldsymbol{\kappa_2 = 5\times10^{-6}}$}}
    & \shortstack{$\kappa_1 = 2.0$\\$\kappa_2 = 5\times10^{-6}$}
    & \shortstack{$\kappa_1 = 5.0$\\$\kappa_2 = 0.0$} \\
    \cmidrule(lr){5-6}
    & & & & RW & No RW & & \\
    \midrule
    Feas.\ Rate (\%)      & 0.69 & 11.06 & 10.52 & \textbf{12.82} & 0.96 & 6.75 & 0.19 \\
    Mean $e$ [NU]         & $2.09\times10^{-3}$ & $5.56\times10^{-4}$ & $5.62\times10^{-4}$ &
                $\mathbf{5.19\times10^{-4}}$ & $8.84\times10^{-4}$ & $7.96\times10^{-4}$ & $1.56\times10^{-3}$ \\
    Mean $\Delta v$ [m/s] & 356.45 & 356.97 & 353.10 & 354.45 & 357.84 & 352.12 & \textbf{339.34} \\
    Mean $\tau_s$ [days]  & 37.39 & 31.11 & 31.39 & \textbf{29.52} & 30.10 & 30.12 & 36.38 \\
    \bottomrule
  \end{tabular}
  \endgroup
  \label{tab:kappa-gamma-comp}
\end{table}
As $\kappa_1$ increases, the focus of the target distribution shifts to generating samples close to Pareto optimality, while less weight is attributed to the feasibility of those samples. 
We can therefore observe a trend of a lower feasibility rate for larger $\kappa_1$, while the mean $\Delta v$ decreases. 
A comparison of columns three and four highlights the influence of $\kappa_2$, with its increase leading to a lower mean shooting time. 
The choice of final parameters as the ones in column four ($\kappa_1=1.0, \kappa_2=5\times10^{-6}$, with reward weighting (RW)) offers the best balance of feasibility and Pareto optimality. 
Although a smaller $\kappa_2$-value increases the mean average shooting time, no significant improvement regarding fullness of the Pareto front for larger shooting times was observed with this adaptation.
Compared to the baseline model, the selected fine-tuned model increases the feasibility ratio by more than an order of magnitude from $0.69\,\%$ to $12.82\,\%$, while also improving $\Delta v$ and $\tau_s$.

In addition to the sample improvement through MCMC, the reward-weighted likelihood optimization is a key factor driving the fine-tuned model’s enhanced performance.
To highlight the effect of the additional reward weighting term in Eq \eqref{eq:updated loss}, Table \ref{tab:kappa-gamma-comp} compares the performance of the selected model trained with reward weighting (RW) to a training run where the reward value is constant across all samples (No RW).
By contributing more weight to higher quality samples during training, the generated samples from the training with reward weighting achieve both lower constraint violations and lower objective values. 
While the model trained without reward weighting still outperforms the baseline model, its feasibility rate is more than an order of magnitude worse than for the model trained with reward weighting. 

\begin{table}[b!]
  \centering
  \caption{Performance metrics based on samples from the fine-tuned diffusion model for $T_{max}=2.741\,\mathrm{N}$ and $T_{max}=4.735\,\mathrm{N}$ in comparison to the baseline model. All metrics are computed based on $100{,}000$ samples with feasibility tolerance $e<10^{-4}\,$ NU.}
  \label{tab:data_ratios_combined}
  \begingroup
  \setlength{\tabcolsep}{4pt}
  \renewcommand{\arraystretch}{1.1}
  \small
  \begin{tabular}{c l c c cccc}
    \toprule
    $T_{max}\,[\mathrm{N}]$ & Metric & Baseline Model & \multicolumn{5}{c}{Fine-tuned Model} \\
    \cmidrule(lr){4-8}
           &        &                & $\eta=0.0$ & $\eta=0.25$ & $\eta=0.5$ & $\eta=0.75$ & $\eta=1.0$ \\
    \midrule
    \multirow{4}{*}{2.741} 
          & Feas.\ Rate (\%)      & 9.79 & 1.03 & 13.77 & 8.94 & \textbf{14.91} & 3.34 \\
          & Mean $e$ [NU]         & $8.06\times 10^{-4}$ & $1.62\times 10^{-3}$ & $5.77\times 10^{-4}$ & $7.06\times 10^{-4}$ & $\mathbf{5.68\times 10^{-4}}$ & $8.43\times 10^{-4}$ \\
          & Mean $\Delta v$ [m/s] & 369.39 & \textbf{345.53} & 369.00 & 367.00 & 371.02 & 359.75 \\
          & Mean $\tau_s$ [days]  & 44.76 & 58.66 & 45.23 & 47.54 & \textbf{43.05} & 59.19 \\
    \midrule
    \multirow{4}{*}{4.735} 
          & Feas.\ Rate (\%)      & 0.69 & \textbf{12.82} & 12.43 & 9.60 & 10.88 & 7.86 \\
          & Mean $e$ [NU]         & $2.09\times 10^{-3}$ & $5.19\times 10^{-4}$ & $\mathbf{5.00\times 10^{-4}}$ & $6.54\times 10^{-4}$ & $5.30\times 10^{-4}$ & $5.55\times 10^{-4}$ \\
          & Mean $\Delta v$ [m/s] & 356.45 & 354.45 & 353.85 & 354.59 & 353.69 & \textbf{353.19} \\
          & Mean $\tau_s$ [days]  & 37.39 & 29.52 & 29.91 & \textbf{29.51} & 30.40 & 31.64 \\
    \bottomrule
  \end{tabular}
  \endgroup
\end{table}
While training the model exclusively on newly generated data significantly improves performance at $T_{max}=4.735\,\mathrm{N}$, it leads to overfitting and degraded performance at other thrust levels. 
This behavior is illustrated in the upper part of Table \ref{tab:data_ratios_combined}, which reports performance metrics at $T_{max}=2.741\,\mathrm{N}$.
At this lower thrust level, the model fine-tuned solely on new data ($\eta=0$) performs much worse than the baseline, with the feasibility rate dropping from $9.79\,\%$ to $1.03\,\%$. 
However, this performance degradation can be mitigated by incorporating a portion of the baseline data into the diffusion model’s loss function, shown as the second term in Eq. \eqref{eq:updated loss}.

Table \ref{tab:data_ratios_combined} summarizes the results for both $T_{max}=2.741\,\mathrm{N}$ and $T_{max}=4.735\,\mathrm{N}$ across various data ratios $\eta$.
In fact, including some baseline data during fine-tuning can improve the performance at $T_{max}=2.741$ N for certain ratios (namely $\eta=0.25$ and $\eta=0.75$) relative to the baseline, even though no new training data at this thrust level was used.
These results suggest that mixing newly generated data with baseline data not only improves the model’s fit to the target distribution at the fine-tuned thrust level, but also yields a better global fit across other thrust magnitudes. 
The performance for $T_{max}=4.735\,\mathrm{N}$ declines as more baseline training data is included (in comparison to $\eta=0.0$).
Overall, $\eta=0.25$ provides a favorable trade-off: it achieves nearly the same performance as $\eta=0$ at $T_{max}=4.735\,\mathrm{N}$ while boosting the feasibility rate at $T_{max}=2.741\,\mathrm{N}$ by over $40\,\%$ (from $9.79\,\%$ to $13.77\,\%$) relative to the baseline.

\subsection{Problem 2: New Model for Titan DRO Transfer}
\begin{table}[b!]
\centering
\caption{Problem parameters for Titan DRO transfer.} 
\setlength{\tabcolsep}{8pt} 
\begin{tabular}{lclc} 
\toprule
\multicolumn{2}{l}{\textbf{Trajectory parameters}} & \multicolumn{2}{c}{} \\
\midrule
\multicolumn{2}{l}{Initial state $[\boldsymbol{r}_0^T,\boldsymbol{v}_0^T]$ [NU]} & \multicolumn{2}{c}{$[1.0758, 0.0, 0.0, 0.0, -0.1684, 0.0]$}\\
\multicolumn{2}{l}{Terminal state $[\boldsymbol{r}_f^T,\boldsymbol{v}_f^T]$ [NU]} & \multicolumn{2}{c}{$[1.0304, 0.0, 0.0, 0.0, -0.1248, 0.0]$}\\
\multicolumn{2}{l}{Orbital period target DRO $\mathcal{T}_f$ [TU]} & \multicolumn{2}{c}{$4.6558$}\\
\multicolumn{2}{l}{Max. shooting time $\tau_{s,.max}$ [TU]} & \multicolumn{2}{c}{$90$}\\
\midrule
\multicolumn{2}{l}{\textbf{Spacecraft parameters}} & \multicolumn{2}{l}{\textbf{Natural units (Saturn-Titan)}} \\
\midrule
Initial mass $m_0$ [kg] & 25,000 & Distance unit [km] & 1,221,870 \\
Fuel mass [kg] & 15,000 & Time unit [s] & 219,277.51 \\
Dry mass [kg] & 10,000 & Mass unit [kg] & $5.685 \times 10^{26}$ \\
Specific impulse $I_{sp}$ [s] & 2,987 & Mass parameter $\mu_{ST}$ & $2.366 \times 10^{-4}$\\
Thrust $T_{max}$ [N] & $0.4500$ & & \\
\bottomrule
\end{tabular}
\label{tab:problem parameters Titan DRO}
\end{table}

\begin{figure}[t!]
  \centering
  %
  \begin{subfigure}[b]{0.45\textwidth}  
    \centering
    \includegraphics[height=5.9cm]{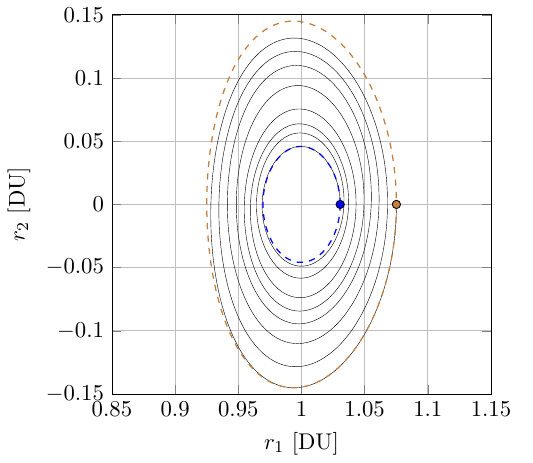}
  \end{subfigure}
  \hfill
  \begin{subfigure}[b]{0.54\textwidth}  
    \centering
    \includegraphics[height=5.9cm]{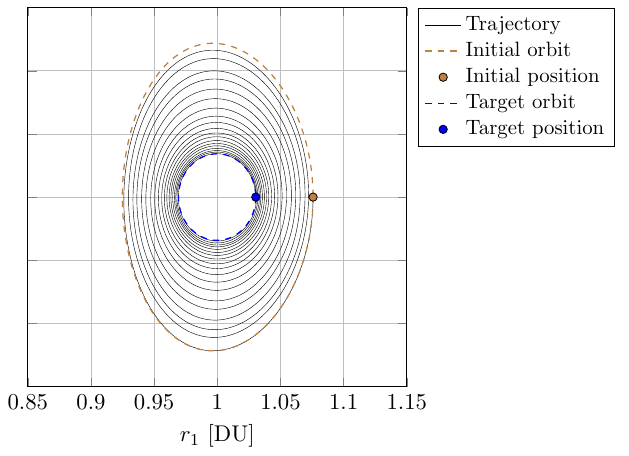}
  \end{subfigure}

  \caption{Example trajectories for the Europa DRO transfer (left) and Titan DRO transfer (right).}
  \label{fig:traj_examples}
\end{figure}

While the first problem highlights the effectiveness of our framework, it presents a relatively simple scenario, since the baseline model was already configured to generate samples for the correct transfer.
To introduce greater complexity, we now fine-tune the baseline model to generate high-quality samples for a new transfer with a different CR3BP mass parameter. 
In particular, we solve a related transfer in the Saturn-Titan system, which has a mass parameter $\mu$ nearly ten times larger than that of Jupiter–Europa.
All problem parameters for this new transfer are presented in Table \ref{tab:problem parameters Titan DRO}.
The baseline model is again trained on the Europa DRO transfer from the previous section, with the initial samples now drawn for a thrust magnitude of $T_{max}=4.984\,\mathrm{N}$.
To preserve the same non-dimensional conditions, the thrust magnitude and specific impulse are set to the natural-unit values used for Europa, giving different SI values.
By choosing the initial and terminal positions of the Saturn DRO such that their distances to the secondary body match those of the Europa DRO transfer, we ensure a comparable problem setup. 
The corresponding velocities are selected to ensure the orbits are closed.
A comparison of two example trajectories is presented in Figure \ref{fig:traj_examples}.
Although the trajectory solutions have a similar structure due to solving the same type of transfer and with the same underlying dynamical model, the change in the model parameter for the mass of the systems induces a noticeably change in the density and number of revolutions required to complete a feasible trajectory. 
For the Titan DRO the energy gap between the departure and arrival orbits is $4.11\times10^{-3}$ NU - noticeably larger than the $2.38\times10^{-3}$ NU gap for the Europa DRO transfer.
Bridging this larger gap requires the spacecraft to complete more revolutions around Titan, with the mean number of revolutions in the final dataset being $23.7$ in comparison to $10.3$ for the Europa DRO transfer.

We first present results obtained by directly targeting the Titan DRO transfer using our MCMC algorithm.
Since this approach fails to produce a complete Pareto front, we then present results from a homotopy method in which the mass parameter $\mu$ is gradually increased.

\subsubsection{Direct MCMC}
\begin{figure}[b!]
	\centering\includegraphics[width=\textwidth]{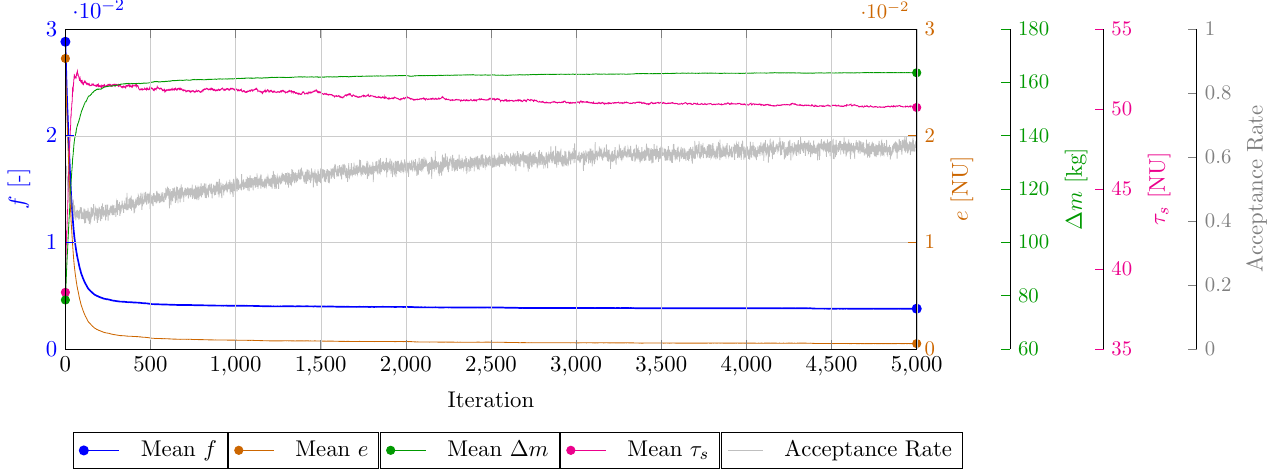}
	\caption{Mean values of the objective function and its three components during direct MCMC for the Titan DRO problem.}
	\label{fig:MCMC_dir_iter_plot}
\end{figure}
\begin{figure}[t!]
  \centering
  \begin{minipage}[t]{0.50\textwidth}
    \vspace{0pt}
    \centering
    \includegraphics[width=\linewidth]{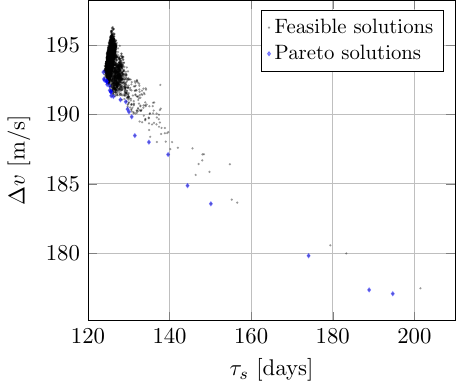}
    \caption{Feasible MCMC samples from the direct approach in the $\Delta v$-$\tau_s$ plane.}
    \label{fig:MCMC_dir_sat_dv_TOF}
  \end{minipage}%
  \hfill
  \begin{minipage}[t]{0.45\textwidth}
    \vspace{0pt}
    \centering
    \captionof{table}{MCMC parameters for the Titan Dro transfer with direct MCMC.}
    \begin{tabular}{@{}ll@{}}
    \toprule
    \textbf{Input Parameters} & \textbf{Value} \\ 
    \midrule
    Objective scaling $\kappa_1$ & $0.5$ \\
    Shooting time scaling $\kappa_2$ & $0.0$ \\
    Proposal standard deviations $\boldsymbol{\sigma}$ & $0.1 \times \boldsymbol{\sigma}_{\text{init}}$ \\
    Scaling factor $\beta$ & $10,000$ \\
    Number of chains $N$ & $1,280$ \\
    Number of iterations $M$ & $5,000$  \\
    Burn-in iterations $M_0$ & $4,950$ \\
    \midrule
  \multicolumn{2}{@{}l}{\textbf{Resulting Metrics}}\\
  Total function evaluations $MN$ & $6,400,000 $\\
  Number of final samples & $40,556$\\
  Runtime [CPU-hours] & 287\\
  \bottomrule
  \end{tabular}
    \label{tab:mcmc_params_direct}
  \end{minipage}
\end{figure}

With the direct approach, we run the random-walk Metropolis sampler for a total of 287 CPU-hours distributed across 32 cores. 
The corresponding parameters are listed in Table \ref{tab:mcmc_params_direct}.
As the initial samples come from the Europa-DRO transfer, they are far from being feasible as emphasized by the large initial mean constraint violation in Figure \ref{fig:MCMC_dir_iter_plot} (contrast Figure \ref{fig:MCMC_iter_europa}).
Moving these samples towards a lower objective function requires the algorithm to cover greater distances in the costate space, which motivates the larger proposal covariance in comparison to the first problem (increased standard deviation from $0.01\boldsymbol{\sigma}_{\text{init}}$ to $0.1\boldsymbol{\sigma}_{\text{init}}$).
The chain quickly reduces both the objective and the constraint violation. In the early iterations, this improvement requires extra fuel and longer shooting times to span the larger energy gap.
Once feasibility improves, those secondary costs dominate the objective, so the mean shooting time decreases and the mean fuel consumption plateaus.
After burn-in, $40,556$ unique samples remain, with a final mean constraint violation of $2.71\times10^{-4}$.

The algorithm yields a large amount of feasible samples, however no full Pareto front is uncovered, as shown by the $\Delta v-\tau_s$ plot in Figure \ref{fig:MCMC_dir_sat_dv_TOF}.
After filtering for $e<10^{-4}$ NU, the feasibility rate is $20.32\,\%$.
Most of these samples are clustered in a region of very short shooting times, with many thrusting at maximum throttle during almost the entire shooting time.
Only very few samples occupy low-$\Delta v$ regions.
This cluster is likely easier for the random-walk Metropolis search to reach - either because it lies closer to the initial states, or spans a larger volume in costate space. 
That hypothesis also explains the trend of an increasing acceptance rate in Figure \ref{fig:MCMC_dir_iter_plot} as the diffusion towards this region that is easier to explore results in more samples from the proposal being accepted.
Although  $\kappa_2=0$ gives no direct incentive to shorten the shooting time, the chain still drifts toward these quick, high-thrust solutions.
Increasing $\kappa_1$ does not lead to a substantial Pareto front improvement and mainly lowers the feasibility rate.

\subsubsection{Homotopy MCMC}
As the direct MCMC approach produces unsatisfactory results, we aim to instead steer the solutions towards Pareto optimality, by solving the problem in multiple small steps. 
We introduce the dimensionless homotopy parameter \(h\) as
\[
h(\mu)=\frac{\mu-\mu_{\text{JE}}}{\mu_{\text{ST}}-\mu_{\text{JE}}},
\quad \mu_{\text{JE}}=2.525\times10^{-5},\;
\mu_{\text{ST}}=2.366\times10^{-4},
\]
so that \(h=0\) corresponds to the Jupiter–Europa system and \(h=1\) to the Saturn–Titan system.
We create a sequence of artificial CR3BP systems with intermediate $\mu$-values and scale each system’s natural units linearly between the two end cases.  
For every intermediate system, we compute closed departure and arrival DROs whose radii match those of the original Europa transfer, while holding specific impulse and thrust magnitude constant in each system’s natural units. 

\begin{table}[t!]
  \centering
  \caption{MCMC parameters for the Titan DRO transfer with homotopy MCMC. Values in brackets indicate 
  adapted parameters for the last $500$ iterations.}
  \label{tab:mcmc_params_homotopy}
  \begin{tabular}{@{}llll@{}}
    \toprule
    \textbf{Input Parameters} & \textbf{Value} & \textbf{Resulting Metrics} & \textbf{Value} \\
    \midrule
    Objective scaling $\kappa_1$ & $1.0$ & Total function evaluations $MN$ & $5{,}760{,}000$ \\
    Shooting time scaling $\kappa_2$ & $1\times10^{-6}$ & Number of final samples & $31{,}400$ \\
    Proposal std.\ devs.\ $\boldsymbol{\sigma}$ & $0.05 \times \boldsymbol{\sigma}_{\text{init}}$ ($0.002 \times \boldsymbol{\sigma}_{\text{init}}$)& Runtime [CPU-hours] & $357$ \\
    Scaling factor $\beta$ & $10{,}000$ ($200{,}000$) & & \\
    Number of chains $N$ & $1{,}920$ & & \\
    Number of iterations $M$ & $3{,}000$ & & \\
    Burn-in iterations $M_0$ & $2{,}470$ & & \\
    \bottomrule
  \end{tabular}
\end{table}
We run the MCMC algorithm for a total of $357$ CPU-hours distributed across $96$ cores.
All MCMC parameters for this problem are listed in Table \ref{tab:mcmc_params_homotopy}.
The proposal covariance is chosen to be smaller than for the direct case (decreased standard deviation from $0.1\boldsymbol{\sigma}_{\text{init}}$ to $0.05\boldsymbol{\sigma}_{\text{init}}$), as moving from one artificial system to the next requires smaller steps than going directly from the Jupiter-Europa to the Saturn-Titan system. 
However it is still larger than for the first problem ($0.01\boldsymbol{\sigma}_{\text{init}}$), which did not involve switching to a new transfer scenario. 
The MCMC algorithm runs for $250$ iterations for each intermediate system, with an additional $500$ iterations with decreased proposal covariance in the final system.
Combined with an increased $\beta$ (values shown in brackets in Table \ref{tab:mcmc_params_homotopy}), these final smaller steps help the algorithm to better resolve local minima. 
Further increasing the number of samples or iterations for this problem only gives marginal improvements.
After discarding the $10\,\%$ samples with the lowest objective value, $31,400$ samples are used to fine-tune the baseline diffusion model using reward-weighted likelihood optimization.
Due to the discrepancy between the solution distribution for this problem and the distribution learned by the baseline model, we use none of the baseline data for training the new model ($\eta=0.0$).

\begin{figure}[b!]
\centering\includegraphics[width=\textwidth]{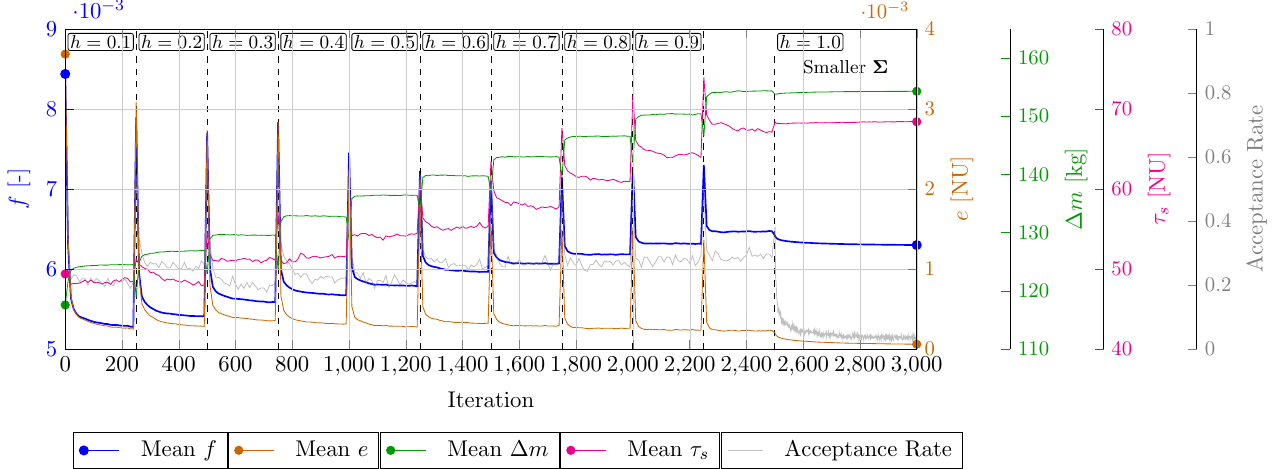}
	\caption{Mean values of the objective function and its three components during Homotopy MCMC for the Titan DRO problem.}
	\label{fig:MCMC_iter_sat_homotopy}
\end{figure}

Across the intermediate homotopy stages, both the mean objective and mean constraint violation drop substantially within each MCMC run. 
Recurring spikes every 250 iterations in Figure \ref{fig:MCMC_iter_sat_homotopy} mark transitions to the next system (new $\mu$). 
From stage to stage, average fuel consumption and shooting time rise, reflecting the growing energy gap and additional revolutions needed. 
Yet within each stage, after an initial jump, the shooting time trends downward and fuel consumption remains nearly flat.
An additional decrease in $f$ is achieved through the decreased proposal covariance and increased $\beta$ in the final stage, as the smaller step size drives down the constraint violations. 
While the acceptance rate is in between $20\,\%$ and $40\,\%$ for the first $2,500$ iterations, it drops to below $5\,\%$ during these final iterations. 
This is due to the increased $\beta$ value, which prevents too many samples with worse objective values from being accepted, as exploration of new local minima is not the primary focus during that stage.

\begin{figure}[b!]
  \centering
  \newlength{\triFigH}
  \setlength{\triFigH}{5.0cm} 

  \begin{minipage}[c]{0.34\textwidth}
    \centering
    \includegraphics[width=\linewidth,height=\triFigH,keepaspectratio]{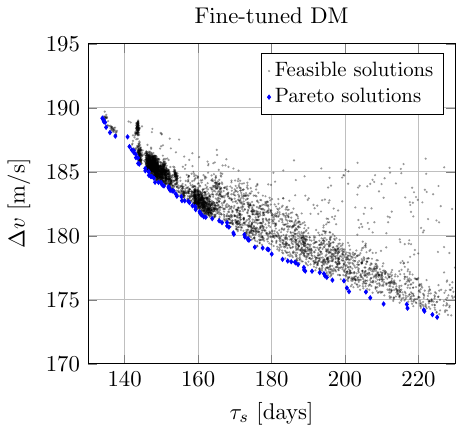}
  \end{minipage}\hfill
  \begin{minipage}[c]{0.31\textwidth}
    \centering
    \includegraphics[width=\linewidth,height=\triFigH,keepaspectratio]{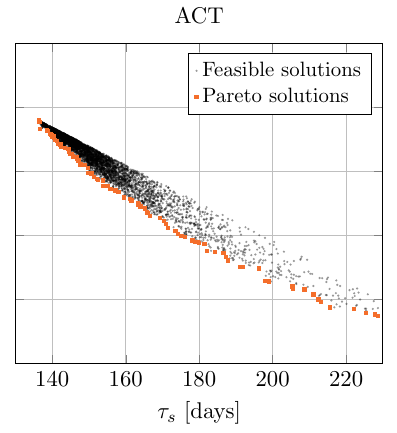}
  \end{minipage}\hfill
  \begin{minipage}[c]{0.34\textwidth}
    \centering
\includegraphics[width=\linewidth,height=\triFigH,keepaspectratio]{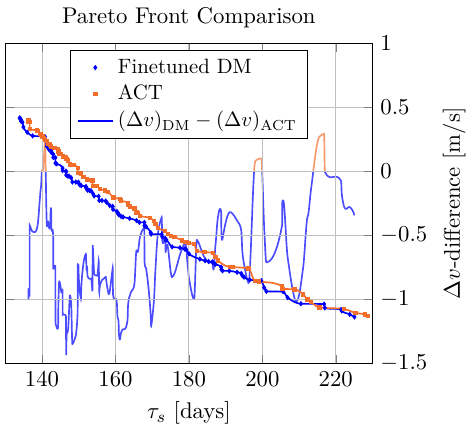}
  \end{minipage}

  \caption{Feasible samples (tolerance $e<5\times10^{-5}$ NU) from the fine-tuned DM compared to feasible samples generated using ACT, shown in the $\Delta v$-$\tau_s$ plane. The presented $5,000$ DM solutions where randomly selected, whereas the $5,000$ ACT solutions are selected based on the best objective values out of a run with $5,760,000$ initializations. The plot on the right compares the Pareto fronts by fitting a piecewise cubic spline to each front and plotting the difference on a separate axis (color of the curve corresponds to the better data in a certain $\tau_s$ range).}
  \label{fig:dv-TOF_saturn_comp}
\end{figure}

Feasible samples generated by the new diffusion model reveal a Pareto front structure in the $\Delta v$-$\tau_s$ plane for this problem, as shown on the left of Figure \ref{fig:dv-TOF_saturn_comp}.
We filter the solutions based on feasibility $e<5\times10^{-5}$ NU, which, when converted to SI units, results in a similar position tolerance as the $10^{-4}$ NU used in the Jupiter-Europa problem.
While that tolerance is still coarse for mission design it is sufficient for this low-fidelity global search.
The left plot of Figure \ref{fig:dv-TOF_saturn_comp} demonstrates the higher quality of feasible homotopy fine-tuning samples in comparison to the corresponding Figure \ref{fig:MCMC_dir_sat_dv_TOF} for the direct MCMC approach.
This result demonstrates that, by seeding an MCMC run with samples from a related problem, the diffusion model can learn a distribution of Pareto optimal solutions without requiring a separate dedicate and focused search for new solver data.
Although we cannot formally certify Pareto optimality, the front likely approximates the true Pareto set. 
Its lower smoothness and density relative to the Europa case (Figure \ref{fig:eur_pareto}) indicate that some Pareto optimal solutions may still be missing.

Our framework yields higher-quality solutions compare to an adjoint-control transformation (ACT) approach.
For a fair comparison, we evaluate $5,760,000$ costate initializations generated with ACT, matching the total number of function evaluations in the MCMC algorithm.
We filter these initializations for feasibility using the preliminary screening algorithm. 
From the feasible set, we select the top $5,000$ in terms of the linear combination of the two objectives. 
Figure \ref{fig:dv-TOF_saturn_comp} compares these solutions and their Pareto optimal subset to an equally large, but randomly selected solution set generated by the fine-tuned DM.
Although the ACT approach also generates solutions close to Pareto optimality, our framework targets Pareto optimality more explicitly.
The better solution quality is highlighted by the direct comparison of the Pareto fronts.

In addition to improved solution quality, our framework offers two key advantages over ACT. First, finding good sampling ranges for the in-plane thrust angle $\alpha$, the switching function $S$ and their time derivatives is tedious and requires multiple test runs. 
These ranges must be specified in order to uniformly sample the control parameters and evaluate the ACT mapping $(\alpha,\dot{\alpha},S,\dot{S})\rightarrow(\boldsymbol{\lambda}_r,\boldsymbol{\lambda}_v)$.
The presented results were only achieved by hand-tuning the control parameter ranges (final choice: $\alpha\in\pi+[-0.0036,0.009]$, $\dot{\alpha}\in[-0.01,0.01]$, $S\in[0.0,0.035]$ and $\dot{S}\in[-0.0012,0.0]$) to target Pareto optimality.
Second, once the fine-tuned DM is trained, we can sample from its learned distribution indefinitely, generating high-quality solutions at a low cost. 

For a quantitative performance comparison, Table \ref{tab:finetuning_results} reports key metrics for the ACT, MCMC, and DM approaches.
The MCMC approach here refers to the final MCMC samples used to train the DM.
Both the MCMC and DM samples outperform the ACT-generated solutions, offering higher feasibility and lower mean $\Delta v$.
Comparing the MCMC and DM columns emphasizes that the final step of training a diffusion model on the MCMC samples is not necessary if only a small amount of high-quality samples is required.
The feasibility rate of samples actually decreases when comparing the final MCMC samples to samples from the diffusion model.
However, the model learns an actual distribution, that can be resampled indefinitely.
This additional exploration step leads to the decreased feasibility ratio.
Through the reward-weighted likelihood maximization, the model focuses more on high-quality samples, with a lower mean objective value.
Due to the choice of weighting parameters in the objective function, a lower fuel consumption is prioritized over minimizing the shooting time for this problem, which explains the decreased mean $\Delta v$ and increased mean $\tau_s$ for the diffusion model samples.
The higher variance for both variables indicates in the DM case shows that the fine-tuning process increases sample diversity.
\begin{table}[b!]
  \centering
  \captionof{table}{Solutions from MCMC samples and fine-tuned DM samples in comparison to separately generated solutions using an ACT method. Mean and standard deviation (STD) are shown for both objectives; feasibility rate is based on $e<5\times10^{-5}$. Metrics are based on at least $30,000$ samples per method.}
    \label{tab:finetuning_results}
    \begin{tabular}{@{}llll@{}}
      \toprule
      & \textbf{ACT} & \textbf{MCMC} & \textbf{DM} \\
      \midrule
     $\Delta v$ (Mean $\pm$ STD) [m/s] & $186.77\pm2.53$ & $185.65\pm2.97$ & $182.55\pm3.34$ \\
     $\tau_s$ (Mean $\pm$ STD)  [days] & $149.88\pm13.52$  & $149.59\pm13.82$ & $168.49\pm23.24$ \\
     Feasibility rate & 0.20\,\% & 17.34\,\% & 5.55\,\% \\
     \bottomrule
\end{tabular}
\end{table}
\begin{figure}[t!]
  \centering
  \begin{subfigure}[t]{0.32\textwidth}
    \centering
    \includegraphics[width=\linewidth]{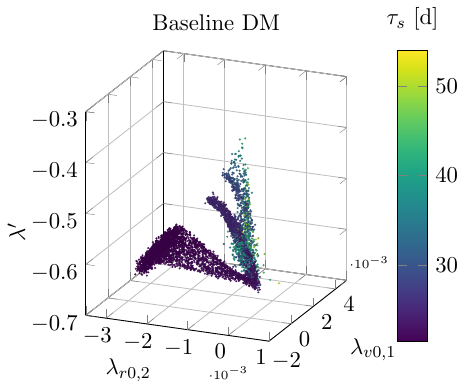}
  \end{subfigure}
  \hfill
  \begin{subfigure}[t]{0.32\textwidth}
    \centering
    \includegraphics[width=\linewidth]{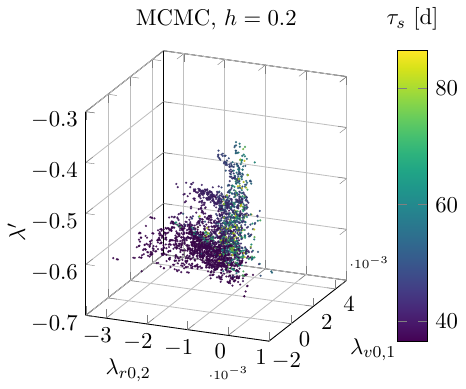}
  \end{subfigure}
  \hfill
  \begin{subfigure}[t]{0.32\textwidth}
    \centering
    \includegraphics[width=\linewidth]{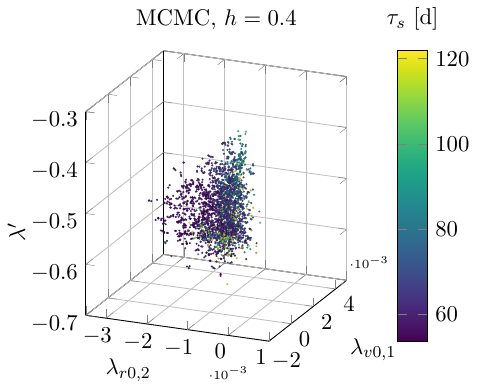}
  \end{subfigure}

  \vspace{0.8em} 

  \begin{subfigure}[t]{0.32\textwidth}
    \centering
    \includegraphics[width=\linewidth]{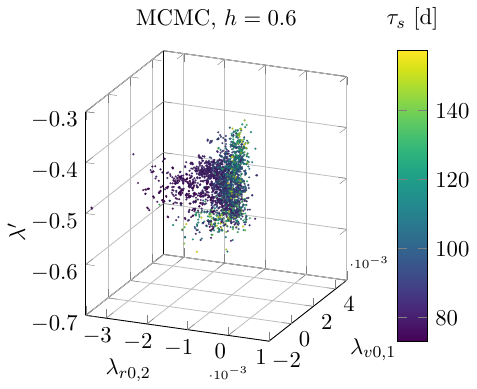}
  \end{subfigure}
  \hfill
  \begin{subfigure}[t]{0.32\textwidth}
    \centering
    \includegraphics[width=\linewidth]{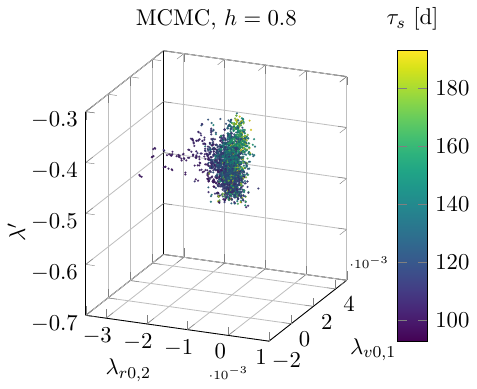}
  \end{subfigure}
  \hfill
  \begin{subfigure}[t]{0.32\textwidth}
    \centering
    \includegraphics[width=\linewidth]{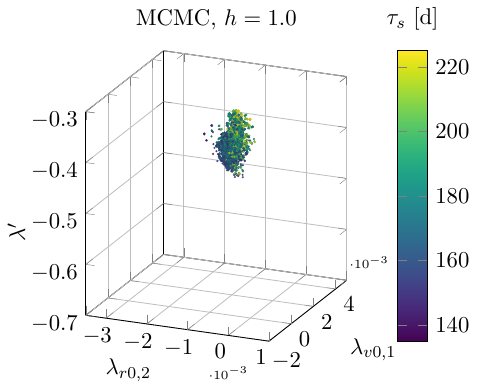}
  \end{subfigure}

  \caption{Comparison of MCMC samples in the costate space for different $h$ factors.}
  \label{fig:mcmc_grid}
\end{figure}

Figure \ref{fig:mcmc_grid} demonstrates how the homotopy approach facilitates the global search process.
It traces the evolution of the sample cloud in costate space during the intermediate MCMC stages.
Starting from the initial hypersurface structure captured by the baseline model, the samples diffuse to a new region of the solution space.
The considerable distance of the final samples for $h=1$ to most of the samples from the baseline model explains why a direct MCMC run could not readily discover these solutions.

\section{conclusion}
In this work we presented a framework that combines Markov chain Monte Carlo-enabled self-supervised fine-tuned diffusion modeling to provide Pareto optimal initial costates for indirect low-thrust trajectory optimization problems. 
The new framework is flexible and does not require an independent and dedicated data generation phase.
Based on the convergence theory of Markov chain Monte Carlo, the framework can start from zero knowledge of the problem at hand, but the convergence rate is improved if samples from a previously trained condition diffusion model that is similar to samples from the target distribution are used for the initial states of the Markov chains. 
In this latter case, numerical experiments showed how the algorithm can explicitly target Pareto optimal solutions and increases the feasibility rate by more than an order of magnitude.
Additionally, the framework outperforms state-of-the-art methods in terms of solution quality when targeting Pareto optimality for new problems.
When the initial distribution for the Markov chain is significantly different than the target distribution, a scheduled homotopy approach can be used with fine-tuning to produce a diffusion model for the target distribution in a stable training manner. 
Thus, we overcome the main drawbacks associated previously with using diffusion models to learn distributions of high-quality solutions, completion of Pareto optimal fronts, and generation of distributions for similar optimal control problems.  

The main limitation of the current framework is potentially long run times of the Markov chain Monte Carlo algorithm and a related reduced performance when solutions to the target and baseline (initial) transfers differ too much.
The latter limitation has been initially addressed in this work with the homotopy-based method, but could be further improved for more general problems and with feedback from the MCMC self-supervised fine-tuning diffusion model process. 
Regarding the former issue, current work by the authors is focused on reducing convergence times and addressing more complex trajectory optimization problems by leveraging more advanced Markov chain Monte Carlo variants. 
The investigations include alternative proposal for the Markov chain, including tuning of random-walk type methods such as a pre-Crank–Nicolson MCMC \cite{Beskos:2008.sd.8.3, Cotter:2013.ss.28.3} and the use of non-Gaussian heavy-tailed proposal distributions (e.g., Pareto and Cauchy). 
We are also actively working to incorporate gradient information by replacing the random-walk Metropolis sampler with gradient-based approaches such as the Metropolis-Adjusted Langevin Algorithm and Hamiltonian Monte Carlo. 
We believe it will be interesting to further investigate the ability of the framework by coupling it with homotopy in different parameters and using the resulting intermediate solutions to train a diffusion model conditioned on the homotopy parameter.

\section{Acknowledgment}
The authors would like to thank Anjian Li for providing the foundational generative learning framework, which served as the basis for our approach, and informative discussions on diffusion modeling. 
Simulations were performed on computational resources managed and supported by Princeton Research Computing, a consortium of groups including the Princeton Institute for Computational Science and Engineering (PICSciE) and the Office of Information Technology’s High-Performance Computing Center and Visualization Laboratory at Princeton University.
The authors would also like to acknowledge partial support for this effort from a Princeton University School of Engineering and Applied Sciences internal seed grant award and gratefully acknowledge partial financial support from the Princeton Laboratory for Artificial Intelligence’s [PLI/AI\textsuperscript{2}/NAM] initiative.

\section{Conflict of Interest}
On behalf of all authors, the corresponding author states that there is no conflict of interest.


\bibliographystyle{AAS_publication}   
\bibliography{references}   

\begin{thebibliography}{10}

\bibitem{li2023amortizedglobalsearchtrajectory}
A.~Li, A.~Sinha, and R.~Beeson, ``Amortized Global Search for Efficient Preliminary Trajectory Design with Deep Generative Models,''  arXiv preprint \url{https://arxiv.org/abs/2308.03960}, 2023.
\newblock \href{https://doi.org/10.48550/arXiv.2308.03960}{10.48550/arXiv.2308.03960}.

\bibitem{beeson2024globalsearchoptimalspacecraft}
R.~Beeson, A.~Li, and A.~Sinha, ``Global Search of Optimal Spacecraft Trajectories using Amortization and Deep Generative Models,''  arXiv preprint \url{https://arxiv.org/abs/2412.20023}, 2024.
\newblock \href{https://doi.org/10.48550/arXiv.2412.20023}{10.48550/arXiv.2412.20023}.

\bibitem{Graebner.1032024}
J.~Graebner, A.~Li, A.~Sinha, and R.~Beeson, ``Learning Optimal Control and Dynamical Structure of Global Trajectory Search Problems with Diffusion Models,''  arXiv preprint \url{https://arxiv.org/abs/2410.02976}, 2024.
\newblock \href{https://doi.org/10.48550/arXiv.2410.02976}{10.48550/arXiv.2410.02976}.

\bibitem{doi:10.2514/6.2012-4517}
J.~Englander, B.~Conway, and T.~Williams, ``Automated Interplanetary Trajectory Planning,''  {\em AIAA/AAS Astrodynamics Specialist Conference}, Minneapolis, MN, USA, August 2012, 10.2514/6.2012-4517.

\bibitem{Yam.2011}
C.~H. Yam, D.~D. Lorenzo, and D.~Izzo, ``Low-thrust trajectory design as a constrained global optimization problem,''  {\em Proceedings of the Institution of Mechanical Engineers, Part G: Journal of Aerospace Engineering}, Vol.~225, No.~11, 2011, pp.~1243--1251, 10.1177/0954410011401686.

\bibitem{Wales.1997}
D.~J. Wales and J.~P.~K. Doye, ``Global Optimization by Basin-Hopping and the Lowest Energy Structures of Lennard-Jones Clusters Containing up to 110 Atoms,''  {\em The Journal of Physical Chemistry A}, Vol.~101, No.~28, 1997, pp.~5111--5116, 10.1021/jp970984n.

\bibitem{li2024diffusolvediffusionbasedsolvernonconvex}
A.~Li, Z.~Ding, A.~B. Dieng, and R.~Beeson, ``DiffuSolve: Diffusion-based Solver for Non-convex Trajectory Optimization,''  arXiv preprint \url{https://arxiv.org/abs/2403.05571}, 2024.
\newblock \href{https://doi.org/10.48550/arXiv.2403.05571}{10.48550/arXiv.2403.05571}.

\bibitem{Li.412025}
A.~Li and R.~Beeson, ``Aligning Diffusion Model with Problem Constraints for Trajectory Optimization,''  arXiv preprint \url{https://arxiv.org/abs/2504.00342}, 2025.
\newblock \href{https://doi.org/10.48550/arXiv.2504.00342}{10.48550/arXiv.2504.00342}.

\bibitem{Graebner.1132025}
J.~Graebner and R.~Beeson, ``Global Search for Optimal Low Thrust Spacecraft Trajectories using Diffusion Models and the Indirect Method,''  arXiv preprint \url{https://arxiv.org/abs/2501.07005}, 2025.
\newblock \href{https://doi.org/10.48550/arXiv.2501.07005}{10.48550/arXiv.2501.07005}.

\bibitem{Betts.1998}
J.~T. Betts, ``Survey of Numerical Methods for Trajectory Optimization,''  {\em Journal of Guidance, Control, and Dynamics}, Vol.~21, No.~2, 1998, pp.~193--207, 10.2514/2.4231.

\bibitem{Lee.2232023}
K.~Lee, H.~Liu, M.~Ryu, O.~Watkins, Y.~Du, C.~Boutilier, P.~Abbeel, M.~Ghavamzadeh, and S.~S. Gu, ``Aligning Text-to-Image Models using Human Feedback,''  arXiv preprint \url{https://arxiv.org/abs/2302.12192}, 2023.
\newblock \href{https://doi.org/10.48550/arXiv.2302.12192}{10.48550/arXiv.2302.12192}.

\bibitem{Kirk.2004}
D.~E. Kirk, {\em Optimal control theory: An introduction}.
\newblock Dover Books on Electrical Engineering Ser, Mineola, N.Y: {Dover Publications}, first published in 2004, unabridged republication of the thirteenth printing~ed., 2004.

\bibitem{Liberzon.2012}
D.~Liberzon, {\em Calculus of variations and optimal control theory: A concise introduction}.
\newblock Princeton, NJ and Woodstock: {Princeton University Pres}, 2012, 10.1515/9781400842643.

\bibitem{Rutherfobd.1964}
D.~E. Rutherfobd, ``Optimal Trajectories For Space Navigation. By D. F. Lawden. Pp. viii, 126. 21s. net. 1963. (Butterworth and Co.),''  {\em The Mathematical Gazette}, Vol.~48, No.~366, 1964, pp.~478--479, 10.2307/3611765.

\bibitem{Pontryagin.2018}
L.~S. Pontryagin, {\em Mathematical Theory of Optimal Processes}.
\newblock Routledge, 2018, 10.1201/9780203749319.

\bibitem{Conway.2010}
B.~A. Conway, {\em Spacecraft Trajectory Optimization}, Vol.~v.29 of {\em Cambridge Aerospace Series}.
\newblock New York: {Cambridge University Press}, 2010.

\bibitem{SohlDickstein.3122015}
J.~Sohl-Dickstein, E.~A. Weiss, N.~Maheswaranathan, and S.~Ganguli, ``Deep Unsupervised Learning using Nonequilibrium Thermodynamics,''  arXiv preprint \url{https://arxiv.org/abs/1503.03585}, 2015.
\newblock \href{https://doi.org/10.48550/arXiv.1503.03585}{10.48550/arXiv.1503.03585}.

\bibitem{Song.7122019}
Y.~Song and S.~Ermon, ``Generative Modeling by Estimating Gradients of the Data Distribution,''  arXiv preprint \url{https://arxiv.org/abs/1907.05600}, 2019.
\newblock \href{https://doi.org/10.48550/arXiv.1907.05600}{10.48550/arXiv.1907.05600}.

\bibitem{Ho.6192020}
J.~Ho, A.~Jain, and P.~Abbeel, ``Denoising Diffusion Probabilistic Models,''  arXiv preprint \url{https://arxiv.org/abs/2006.11239}, 2020.
\newblock \href{https://doi.org/10.48550/arXiv.2006.11239}{10.48550/arXiv.2006.11239}.

\bibitem{Wang.8122022}
Z.~Wang, J.~J. Hunt, and M.~Zhou, ``Diffusion Policies as an Expressive Policy Class for Offline Reinforcement Learning,''  arXiv preprint \url{https://arxiv.org/abs/2208.06193}, 2022.
\newblock \href{https://doi.org/10.48550/arXiv.2208.06193}{10.48550/arXiv.2208.06193}.

\bibitem{Ding.252024}
Z.~Ding, A.~Zhang, Y.~Tian, and Q.~Zheng, ``Diffusion World Model: Future Modeling Beyond Step-by-Step Rollout for Offline Reinforcement Learning,''  arXiv preprint \url{https://arxiv.org/abs/2402.03570}, 2024.
\newblock \href{https://doi.org/10.48550/arXiv.2402.03570}{10.48550/arXiv.2402.03570}.

\bibitem{Tevet.9292022}
G.~Tevet, S.~Raab, B.~Gordon, Y.~Shafir, D.~Cohen-Or, and A.~H. Bermano, ``Human Motion Diffusion Model,''  arXiv preprint \url{https://arxiv.org/abs/2209.14916}, 2022.
\newblock \href{https://doi.org/10.48550/arXiv.2209.14916}{10.48550/arXiv.2209.14916}.

\bibitem{Ronneberger.5182015}
O.~Ronneberger, P.~Fischer, and T.~Brox, ``U-Net: Convolutional Networks for Biomedical Image Segmentation,''  arXiv preprint \url{https://arxiv.org/abs/1505.04597}, 2015.
\newblock \href{https://doi.org/10.48550/arXiv.1505.04597}{10.48550/arXiv.1505.04597}.

\bibitem{Ouyang.342022}
L.~Ouyang, J.~Wu, X.~Jiang, D.~Almeida, C.~L. Wainwright, P.~Mishkin, C.~Zhang, S.~Agarwal, K.~Slama, A.~Ray, J.~Schulman, J.~Hilton, F.~Kelton, L.~Miller, M.~Simens, A.~Askell, P.~Welinder, P.~Christiano, J.~Leike, and R.~Lowe, ``Training language models to follow instructions with human feedback,''  arXiv preprint \url{https://arxiv.org/abs/2203.02155}, 2022.
\newblock \href{https://doi.org/10.48550/arXiv.2203.02155}{10.48550/arXiv.2203.02155}.

\bibitem{Ziegler.9182019}
D.~M. Ziegler, N.~Stiennon, J.~Wu, T.~B. Brown, A.~Radford, D.~Amodei, P.~Christiano, and G.~Irving, ``Fine-Tuning Language Models from Human Preferences,''  arXiv preprint \url{https://arxiv.org/abs/1909.08593}, 2019.
\newblock \href{https://doi.org/10.48550/arXiv.1909.08593}{10.48550/arXiv.1909.08593}.

\bibitem{Russell.2007}
R.~P. Russell, ``Primer Vector Theory Applied to Global Low-Thrust Trade Studies,''  {\em Journal of Guidance, Control, and Dynamics}, Vol.~30, No.~2, 2007, pp.~460--472, 10.2514/1.22984.

\bibitem{Beeson.Aug.2022}
R.~Beeson, A.~Sinha, B.~Jagannatha, D.~Bunce, and D.~L. Carroll, ``Dynamically Leveraged Automated (N) Multibody Trajectory Optimization (DyLAN),''  {\em AAS/AIAA Space Flight Mechanics Conference. American Astronautical Society. Charlotte, NC}, Aug. 2022.

\bibitem{Bentley.1975}
J.~L. Bentley, ``Multidimensional binary search trees used for associative searching,''  {\em Communications of the ACM}, Vol.~18, No.~9, 1975, pp.~509--517, 10.1145/361002.361007.

\bibitem{Metropolis.1953}
N.~Metropolis, A.~W. Rosenbluth, M.~N. Rosenbluth, A.~H. Teller, and E.~Teller, ``Equation of State Calculations by Fast Computing Machines,''  {\em The Journal of Chemical Physics}, Vol.~21, No.~6, 1953, pp.~1087--1092, 10.1063/1.1699114.

\bibitem{Hastings.1970}
W.~K. Hastings, ``Monte Carlo sampling methods using Markov chains and their applications,''  {\em Biometrika}, Vol.~57, No.~1, 1970, pp.~97--109, 10.1093/biomet/57.1.97.

\bibitem{Douc.2018}
R.~Douc, E.~Moulines, P.~Priouret, and P.~Soulier, {\em Markov Chains}.
\newblock Springer Series in Operations Research and Financial Engineering, Cham: {Springer International Publishing} and {Imprint: Springer}, 1st ed. 2018~ed., 2018.

\bibitem{10.1214/aos/1176325750}
L.~Tierney, ``{Markov Chains for Exploring Posterior Distributions},''  {\em The Annals of Statistics}, Vol.~22, No.~4, 1994, pp.~1701 -- 1728, 10.1214/aos/1176325750.

\bibitem{Brooks.1998}
S.~Brooks, ``Markov chain Monte Carlo method and its application,''  {\em Journal of the Royal Statistical Society: Series D (The Statistician)}, Vol.~47, No.~1, 1998, pp.~69--100, 10.1111/1467-9884.00117.

\bibitem{CharlesJ.Geyer.1992}
{Charles J. Geyer}, ``Practical Markov Chain Monte Carlo,''  {\em Statistical Science}, Vol.~7, No.~4, 1992, pp.~473--483.

\bibitem{Dixon.1981}
L.~C.~W. Dixon and M.~C. Bartholomew-Biggs, ``Adjoint-control transformations for solving practical optimal control problems,''  {\em Optimal Control Applications and Methods}, Vol.~2, No.~4, 1981, pp.~365--381, 10.1002/oca.4660020405.

\bibitem{Beskos:2008.sd.8.3}
A.~Beskos, G.~Roberts, A.~Stuart, and J.~Voss, ``MCMC Methods for Diffusion Bridges,''  {\em Stochastics and Dynamics}, Vol.~8, No.~3, 2008, 10.1142/S0219493708002378.

\bibitem{Cotter:2013.ss.28.3}
S.~L. Cotter, G.~O. Roberts, A.~M. Stuart, and D.~White, ``MCMC Methods for Functions: Modifying Old Algorithms to Make Them Faster,''  {\em Statistical Science}, Vol.~28, No.~3, 2013, pp.~424--446.

\end{thebibliography}

\end{document}